\documentclass[aps,prb,onecolumn,nofootinbib,citeautoscript,10pt]{revtex4-2}

\usepackage{amsmath,amssymb} 
\usepackage{comment}

\usepackage[tight]{subfigure} 

\usepackage[dvipsnames]{xcolor} 
\usepackage[papersize={8.5in,11in}]{geometry}
\usepackage[colorlinks=true]{hyperref}
\hypersetup{
    bookmarks=true,         
    unicode=false,          
    pdftoolbar=true,        
    pdfmenubar=true,        
    pdffitwindow=false,     
    pdfstartview={FitH},    
    pdfkeywords={keyword1} {key2} {key3}, 
    pdfnewwindow=true,      
    colorlinks=true,       
    linkcolor=magenta, 
    citecolor=blue,        
    filecolor=magenta,      
    urlcolor=blue           
} 

\usepackage{dcolumn}
\usepackage{color}
\usepackage{amssymb,amsmath}
\usepackage{tabularx,graphicx}
\usepackage{epstopdf}
\usepackage{latexsym}
\usepackage{colortbl}
\usepackage{psfrag}
\usepackage{bbm,bm,array,physics}
\usepackage{dsfont}
\usepackage{float, mathrsfs}

\def \nn{\nonumber \\}
\def\*#1{\boldsymbol{#1}} 

\begin{document}

\title{Longitudinal magnetoconductivity in multifold semimetals exemplified by pseudospin-1 nodal points}

\author{Ipsita Mandal}
\email{ipsita.mandal@snu.edu.in}

\affiliation{Department of Physics, Shiv Nadar Institution of Eminence (SNIoE), Gautam Buddha Nagar, Uttar Pradesh 201314, India}

\begin{abstract} 
We embark on computing the longitudinal magnetoconductivity within the semiclassical Boltzmann formalism, where an isotropic triple-point semimetal (TSM) is subjected to collinear electric ($\boldsymbol E $) and magnetic ($\boldsymbol B$) fields. Except for the Drude part, the $B$-dependence arises exclusively from topological properties like the Berry curvature and the orbital magnetic moment. We solve the Boltzmann equations exactly in the linear-response regime, applicable in the limit of weak/nonquantising magnetic fields. The novelty of our investigation lies in the consideration of the truly multifold character of the TSMs, where the so-called flat-band (flatness being merely an artefact of linear-order approximations) is made dispersive by incorporating the appropriate quadratic-in-momentum correction in the effective Hamiltonian. It necessitates the consideration of interband scatterings within the same node as well, providing a complex interplay of intraband, interband, intranode, and internode processes, offering an overwhelmingly rich set of possibilities. The exact results are compared with those obtained from a naive relaxation-time approximation.
\end{abstract}

\maketitle
\tableofcontents


\section{Introduction}

Motivated by the ongoing interests in unravelling novel transport properties in three-dimensional (3d) nodal-point semimetals~\cite{burkov11_Weyl, armitage_review, yan17_topological, ips-kush-review, bernevig, grushin-multifold, ips-hermann-review, claudia-multifold}, we undertake the exercise of studying pseudospin-1 triple-point semimetals (TSMs)~\cite{bernevig, ips3by2, ady-spin1, krish-spin1, ips-cd1, tang2017_multiple, grushin-multifold, prb108035428, ips-abs-spin1, claudia-multifold, ips-spin1-ph}, with a quadratic correction that renders the system more realistic \cite{ni2021_giant, claudia-multifold}. TSMs provide a platform for us to formulate and observe some generic physical phenomena connected with quantum anomalies, especially in the context of multifold fermions. In the realm of high-energy physics, chiral anomaly is synonymous with the Adler-Bell-Jackiw anomaly of quantum electrodynamics \cite{adler, bell}. Strikingly, this phenomenon continues to show up even in nonrelativistic settings involving 3d semimetals \cite{chiral_ABJ, hosur-review, son13_chiral}. In particular, chiral anomaly from three-component quasiparticles is intriguing as they have no counterpart in relativistic spinors appearing in high-energy physics. This is because spin-statistics theorem prohibits the existence of fermions carrying an integer-spin value of unity.

The existence of multifold fermions in crystal structures can be understood from the following discussions: 3d semimetals comprise an extremely broad and diverse family, with the degeneracy at the nodal point being given by $(2\, \varsigma + 1)$, where the effective Hamiltonian in the vicinity of a node takes the form of $ \boldsymbol d( \boldsymbol{k} ) \cdot \boldsymbol {\mathcal S } $, with
$\boldsymbol d( \boldsymbol{k} ) = 
\lbrace d_x ( \boldsymbol{k} ) , d_y ( \boldsymbol{k} ) , d_z ( \boldsymbol{k} )  
 \rbrace.$
The symbol $\boldsymbol {\mathcal S } $ stands for the vector operator comprising the components $ \lbrace \mathcal S_x, \mathcal S_y, \mathcal S_z  \rbrace $, representing the three components of the angular-momentum operator in the spin-$\varsigma$ representation of the SO(3) or SU(2) group. Consequently, the $(2\, \varsigma + 1)$-band system is said to carry the quantum numbers of pseudospin-$\varsigma$, with the bands labelled by the $\mathcal S_z$-projections spanning from $- \varsigma $ to $ \varsigma $. The minimal case of twofold degeneracy is represented by the pseudospin-1/2-valued Weyl semimetals (WSMs)~\cite{burkov11_Weyl, armitage_review, yan17_topological}, which are isotropic in the simplest scenarios. Their multifold cousins are captured by the pseudospin-1 TSMs and the pseudospin-3/2 Rarita-Schwinger-Weyl (RSW) semimetal~\cite{bernevig, long, igor, igor2, isobe-fu, tang2017_multiple, ips3by2, ips-cd1, ma2021_observation, ips-magnus, ips-jns, prb108035428, ips_jj_rsw, grushin-multifold, claudia-multifold, ips-rsw-ph, ips-shreya}, for example.

All the systems described above exemplify bandstructures featuring nontrivial topology \cite{xiao_review, sundaram99_wavepacket, graf-Nband} induced by the Berry phases, which show up in the form of topological quantities like the Berry curvature (BC) and the orbital magnetic moment (OMM). These, in turn, source nontrivial features in the response, that can be measured in transport-experiments \cite{timm, ips_rahul_ph_strain,rahul-jpcm, ips-kush-review, claudia-multifold, ips-ruiz, ips-rsw-ph, ips-tilted, ips-shreya}. While our primary focus in this paper is computing longitudinal magnetotransport for TSMs with $ \varsigma = 1$, other examples include intrinsic anomalous-Hall effects~\cite{haldane04_berry,goswami13_axionic, burkov14_anomolous}, planar-Hall coefficients \cite{zhang16_linear, chen16_thermoelectric, nandy_2017_chiral, nandy18_Berry, amit_magneto, ips-floquet,  das20_thermal, das22_nonlinear, pal22a_berry, pal22b_berry, fu22_thermoelectric, araki20_magnetic, mizuta14_contribution, ips_rahul_ph_strain, timm, rahul-jpcm, ips-kush-review, claudia-multifold, ips-ruiz, lei_li-ph, ips-tilted, ips-rsw-ph, ips-shreya, ips-spin1-ph, ips-nl-ph, gusynin06_magneto, staalhammar20_magneto, yadav23_magneto}, Magnus-Hall effect~\cite{papaj_magnus, amit-magnus, ips-magnus}, circular dichroism \cite{ips-cd1, ips_cd}, and circular photogalvanic effect \cite{moore18_optical, guo23_light,kozii, ips_cpge}, to name a few.
The effective low-energy model of a TSM comprises threefold-degenerate band-crossing points, as shown in Fig.~\ref{figdis}, for which $ \boldsymbol  {\mathcal S } $ is written down in a spin-1 representation. The pseudospin-1 quasiparticles can be realised in widely-available systems and have been the focus of numerous studies \cite{optical_lat1, optical_lat2, cold-atom, lv, spin13d1, spin13d2, spin13d3, spin13d4, shen,lan, urban, peng-he, lai, ips3by2, krish-spin1, ips-cd1, bitan-spin1, pal22b_berry, ips-abs-spin1, ips-spin1-ph}. In the literature, they have sometimes been referred to as ``Maxwell fermions'' \cite{cold-atom}, borrowing the terminology from the spin-1 quantum numbers of the photons, which emerge from the Maxwell equations for electromagnetism. Their relevance in transport experiments involving multifold nodal points cannot be overemphasised, where the unmissable signatures of chiral anomaly have been observed via planar-Hall set-ups \cite{claudia-multifold}. In particular, these experimental results and the associated theoretical explanations show that chiral anomaly is a generic phenomenon, not necessarily limited to spin-1/2 or pseudospin-1/2 fermions.

A nonzero BC-profile is accompanied by nonzero BC monopoles at the nodal points \cite{fuchs-review, polash-review}, serving as topological charges, and sourcing the BC flux. These charges can be interpreted as the Chern numbers of the Bloch bands, when we consider the topological properties of the 3d Brillouin zone (BZ) treated as a closed manifold. The sign of the Chern number of a valence band is referred to as the chirality ($\chi$) of the node, endowing a label of \textit{handedness or chirality} to the quasiparticles associated with it. They are called \textit{right-handed} or \textit{left-handed} according to whether $\chi = 1$ or $\chi = -1$. The chirality is the origin of the chiral anomaly in topological nodal points. Of course, a summation of all the Chern numbers over all the nodes in the BZ, carried either by the conduction or the valence bands, must yield zero for a system originating from a lattice. The mathematical proof of this phenomenon is provided by the Nielsen-Ninomiya theorem \cite{nielsen81_no}. The outcome of this fact is that there exist conjugate pairs of nodes in the BZ, carrying $\chi = \pm 1$. Here, we will adhere to the convention of assigning $\chi$ the sign of the Chern numbers of the negative-energy bands, where positive or negative is measured with respect to the band-touching point (taken to set the zero of energy).

For topological semimetals, there exist extensive amounts of theoretical and experimental studies, trying to identify and characterise unique response-coefficients~\cite{zhang16_linear, chen16_thermoelectric, nandy_2017_chiral, nandy18_Berry, amit_magneto, das20_thermal, das22_nonlinear, pal22a_berry, pal22b_berry, fu22_thermoelectric, araki20_magnetic, mizuta14_contribution,timm, onofre, ips_rahul_ph_strain, rahul-jpcm, ips-kush-review, claudia-multifold, ips-ruiz, ips-tilted, ips-internode, ips-nl-ph}, which can reflect the underlying topology of the BZ. Nevertheless, more remains to be explored. For example, in our earlier works, we have computed the weak-magnetic-field-induced response in planar-Hall set-ups, in the limit of relaxation-time approximation (RTA). Here, we aim to perform an exact calculation of the longitudinal conductivity, arising from applying collinear electric ($\boldsymbol   E $) and magnetic ($\boldsymbol B $) fields, by going beyond the RTA. This exercise has been carried out for WSMs \cite{timm, girish2023}, which we will incorporate here for TSMs, after the unphysical representation of a flat-band is remedied by adding a spherically-symmetric quadratic-in-momentum correction \cite{ni2021_giant}. Hence, the flat-band will be promoted to a quadratically-dispersive band, which will now contribute to the net conductivity (as expected in an actual experiment \cite{claudia-multifold}). We would like to point out that, in Ref.~\cite{girish-internode-spin1}, the behaviour of the longitudinal conductivity has been discussed for TSMs with a nondispersive band, where it does not contribute, since quadratic corrections are not taken into account.

The paper is organized as follows: In Sec.~\ref{secmodel}, we describe the explicit form of the low-energy effective Hamiltonian in the vicinity of a TSM node, after adding the appropriate quadratic correction. Sec.~\ref{secboltz} is devoted to deriving the equations leading to the final values of the longitudinal magnetoconductivity. The results are discussed in Sec.~\ref{secres} therein, illustrated by representative plots. For examining the validity of the RTA for this multifold system, in Sec.~\ref{secsig}, we rederive the conductivity using this simplifying assumption and compare it with the results obtained from the exact expressions. Finally, we conclude with a summary and outlook in Sec.~\ref{secsum}. In all our expressions, we will be using the natural units, which means that the reduced Planck's constant ($\hbar $), the speed of light ($c$), and the Boltzmann constant ($k_B $) are each set to unity.
Additionally, the electric charge has no units, with the magnitude of a single electronic charge measuring $e =1$. Although $e$ is equal to one, we will retain it in our expressions just as a matter of book-keeping.

\section{Model}
\label{secmodel}

\begin{figure*}[t]
\includegraphics[width = 0.65 \textwidth]{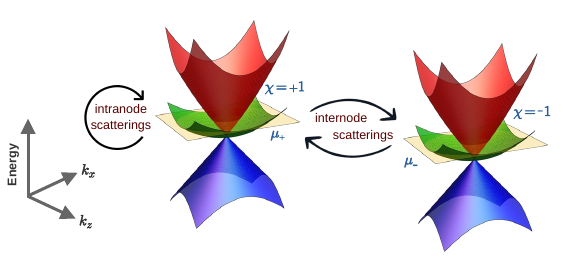}
\caption{\label{figdis}Schematics of the scattering processes between two nodes in a TSM, carrying opposite values of chirality. The values of the chemical potential, represented by the yellow planes, have been tuned to cut the positive-energy bands at each node.}
\end{figure*}

 
Our focus is on TSMs, which exist in materials materials hosting a pair of conjugate nodes with Chern numbers $ \pm 2 $ in the BZ \cite{bernevig, ady-spin1, grushin-multifold}. In particular, we will focus on isotropic nodes, obtained by expanding the $\boldsymbol{k} \cdot \boldsymbol {p}$ Hamiltonian about the threefold-degenerate points, in small $\lbrace k_x, \, k_y , \, k_z \rbrace $, valid in the low-energy limit. We note that, if we decide to expand upto linear-in-momentum terms, we get a flat-band, which is of course an artefact of linear-order approximation and corresponds to an unphysical situation. Therefore, we retain the rotationally-symmetric terms upto quadratic-in-momentum, which is captured by the following effective continuum Hamiltonian \cite{ni2021_giant}:
\begin{align} 
\label{eqham}
\mathcal{H}_\chi ( \boldsymbol k) & = 
\boldsymbol d ( \boldsymbol k) \cdot \boldsymbol  S 
+ c \,v_F \,  k^2  \; \mathbb{I}_{3\times 3} \,,
\quad 
\boldsymbol d( \boldsymbol k) =
v_F \left \lbrace k_x, \,k_y, \, \chi  \, k_z \right \rbrace, 
\quad k =\sqrt{k_x^2 +k_y^2 +k_z^2} \,.
\end{align}
Here, $\boldsymbol  S  = \lbrace S_x, \,S_y, \, S_z \rbrace$ represents the vector comprising the three components of the angular-momentum operator in the spin-$1$ representation of the $SO(3)$ group, $\chi \in \lbrace 1, -1 \rbrace $ denotes the chirality of the node, and $v_F$ is the isotropic Fermi velocity. For our calculations, we have chosen the particular representation where
\begin{align}
& S_x= \frac{1}{\sqrt{2}}
\begin{pmatrix}
	0 & 1  & 0  \\
	1 & 0 & 1  \\
	0 & 1 & 0  
\end{pmatrix} , \quad
S_y=\frac{1}{\sqrt{2}}
\begin{pmatrix}
	0 & -i  & 0  \\
	i & 0 & -i  \\
	0 & i & 0 
\end{pmatrix}, \quad
S_z =
\begin{pmatrix}
	1 & 0 & 0 \\
	0 & 0 & 0 \\
	0 & 0 &-1
\end{pmatrix}.
\end{align}
The model follows from the low-energy $\boldsymbol k \cdot \boldsymbol p$ Hamiltonian near a band-crossing point (taken to be the origin or $ \boldsymbol k = 0$), yielding the Hamiltonian consistent with the symmetries of the little group near the degeneracy point. Refs.~\cite{bernevig, grushin-multifold} expand upto linear order in momentum and the resulting model is a Hamiltonian with a flat-band. The detailed group-theory analysis of the quadratic-corrected model can be found in the ``Supplementary Information'' of Ref.~\cite{ni2021_giant}, where the authors expand about the nodal point upto quadratic order in momentum (they have considered the material CoSi for their experiments). For these quadratic terms, anisotropy is possible \cite{ni2021_giant}, but we retain only the isotropic contributions.

The energy eigenvalues of the Hamiltonian are given by
 \begin{align} 
\label{eigenvalues}
\varepsilon_ s  ({ \boldsymbol k}) = 
 s \,v_F\,  k + c\, v_F \, k^2 \,, \quad
s \in \lbrace -1,0,1 \rbrace \,,
\end{align}
as illustrated in Fig.~\ref{figdis}.
Setting $c$ to zero makes the $s=0$ band morph into a flat (i.e., nondispersive) band, as discussed above. We will consider the $c\, v_F \, k^2$ term to be the subdominant term for the $s = \pm 1$ bands, and choose $c>0$ such that a positive chemical potential ($\mu$) cuts the $s=0$ band.
The group-velocity of the chiral quasiparticles, occupying the band with index $s$, is given by
\begin{align}
{\boldsymbol  v}_ s ( \boldsymbol k) 
\equiv \nabla_{\boldsymbol k} \varepsilon_ s   (\boldsymbol k)
=  \frac{ v_F \left( s + 2 \, c\, k \right) } {  k }  
\left \lbrace k_x,\,k_y, \, k_z \right \rbrace .
\end{align}

Exploiting the spherical symmetry of the problem, we will work in the spherical polar coordinates, such that
\begin{align}
\label{eqcyln} 
k_x =  k \sin \theta  \cos \phi \,, \quad
k_y =   k \sin \theta  \sin \phi \,, \quad k_z = k \cos \theta\,,
\end{align}
where $ k \in [0, \infty )$, $\phi \in [0, 2 \pi )$, and $\theta \in [0, \pi ]$. 
For all the positive-energy bands, a set of normalised orthogonal eigenvectors, $ \lbrace \psi_{\chi, s} (\boldsymbol k) \rbrace $, can be represented as follows:
\begin{align}
\label{eqev}
 \begin{bmatrix}
e^{-2\, i\, \phi } \cos ^2\left(\frac{\theta }{2}\right) &
\,\frac{e^{-i \,\phi } \sin \theta } {\sqrt{2}} &
\,\sin ^2\left(\frac{\theta}{2}\right) 
\end{bmatrix}^{\rm T} &  \quad \text{ for } 
   \{\chi , s\} = \{1, 1\} \,,
\nn \begin{bmatrix}
e^{-2 \,i\, \phi } \sin ^2\left(\frac{\theta }{2}\right) &
\,\frac{ -\, e^{-i\, \phi } \sin \theta } {\sqrt{2}} &
\, \cos ^2\left(\frac{\theta }{2}\right)
\end{bmatrix}^{\rm T} 
& \quad \text{ for }   \{\chi , s\} = \{ -1, 1\} \,,
\nn \begin{bmatrix}
\frac{ -\, e^{-2 \,i\, \phi } \sin \theta } {\sqrt{2}} &
\, e^{-i \, \phi } \cos \theta &
\,\frac{\sin \theta } {\sqrt{2}}
\end{bmatrix}^{\rm T} 
&  \quad \text{ for }
   \{\chi , s\} = \{ \pm 1,  0 \} \,.
\end{align} 
Here, we note that, since the $c \,v_F \,  k^2$ correction accompanies the identity matrix, its addition does not alter the spinor structure of the eigenvectors, when compared to the models considered earlier (for example, in Refs.~\cite{ips-spin1-ph, grushin-multifold}). Thus the topological properties like the BC and the OMM [see Eq.~\eqref{eqomm} below] also remain the same.

\section{Conductivity}
\label{secboltz}

In this section, we will outline the methodology to compute the conductivity using the semiclassical Boltzmann formalism \cite{mermin, sundaram99_wavepacket, li2023_planar, ips-kush-review, ips-rsw-ph, ips-shreya, timm}.

\subsection{Relevant topological quantities}

We discuss here the vectors given by the Berry curvature (BC) and the orbital magnetic moment (OMM), which will affect the linear response that we are set out to compute. For the band with index $s$, these are expressed by the generic formulae of \cite{xiao_review,xiao07_valley}
\begin{align} 
\label{eqomm}
& {\boldsymbol \Omega}_{ \chi, s }( \boldsymbol k)  = 
    i  \left[ \nabla_{ \boldsymbol k}  \psi_{ \chi, s }({ \boldsymbol k}) \right ]^\dagger 
    \cross  \left [ \nabla_{ \boldsymbol k}  \psi_{ \chi, s }({ \boldsymbol k}) \right ]
\text{and } 
{\boldsymbol {m}}_{ \chi, s } ( \boldsymbol k) = 
\frac{  -\,i \, e} {2 } \,
 \left[ \nabla_{ \boldsymbol k} \psi_{ \chi, s } ({ \boldsymbol k}) \right ]^\dagger 
 \cross
\Big [
\left \lbrace \mathcal{H}_\chi({ \boldsymbol k}) -\varepsilon_ s
({ \boldsymbol k}) 
\right \rbrace
\left \lbrace \boldsymbol \nabla_{ \boldsymbol k} \psi_{ \chi, s }({ \boldsymbol k})
 \right \rbrace \Big  ] ,
\end{align}
respectively. Here, $ \lbrace |  \psi_{ \chi, s }({ \boldsymbol k}) \rangle  \rbrace $ is the set of normalized eigenvectors for the parent Hamiltonian. On evaluating the expressions in Eq.~\eqref{eqomm} using  $\mathcal{H}_\chi ( \boldsymbol k)$, we get
\begin{align}
& \boldsymbol \Omega_{\chi,s} ({ \boldsymbol k}) = 
 \frac{ -\, \chi \, s  } { k^3 }
 \left \lbrace k_x, \, k_y, \,  k_z \right \rbrace , \quad
 {\boldsymbol {m}}_{\chi, s} ({ \boldsymbol k}) 
=   \frac{ -\, \chi \, e\, v_F \, \mathcal{G}_s} {2 \, k^2 } 
 \left \lbrace k_x, \, k_y, \,  k_z \right \rbrace,
 \text{ where } 
 \mathcal{G}_ s =\begin{cases}
 1 & \text{ for } s=\pm 1 \\
 2 & \text{ for } s=0
 \end{cases} \, .
\end{align}
We note that, for the $s=0$ band, although the BC is identically zero, the OMM is nonzero, turning out to be twice the OMM of the $s =\pm1 $ bands. Therefore, all the bands are endowed with nontrivial topological properties.
It is easy to verify from the $\boldsymbol \Omega_{\chi, s} ({ \boldsymbol k})$-expressions that the node has a net Chern number of $2\, \chi $, contributed by the $s =\pm 1$ bands.

A nonzero BC modifies the phase-space volume element for the particles occupying a Bloch band via the factor of $
\left [{\mathcal D}_{ \chi, s }  (\boldsymbol k)\right]^{-1}$, where
\begin{align}
{\mathcal D}_{ \chi, s }  (\boldsymbol k) = \left [1 
+ e \,  \left \lbrace 
{\boldsymbol B} \cdot \boldsymbol{\Omega }_{ \chi, s }  (\boldsymbol k)
\right \rbrace  \right ]^{-1}.
\end{align}
On the other hand, the presence of a nonzero OMM causes a Zeeman-like correction to the bare dispersion \cite{xiao_review}, leading to a net effective value of
\begin{align}
\label{eqmodi}
& \xi_{ \chi, s } (\boldsymbol k) 
= \varepsilon_ s  (\boldsymbol k) + \varepsilon_{\chi, s}^{ (m) }  (\boldsymbol k) \, ,
\quad 
\varepsilon_{\chi, s}^{(m)}   (\boldsymbol k) 
= - \,{\boldsymbol B} \cdot \boldsymbol{m }_{\chi, s}  (\boldsymbol k) \,.
\end{align}
This, in turn, modifies the group-velocity as
\begin{align}
{\boldsymbol   w}_{ \chi, s } ({\boldsymbol k} ) \equiv 
 \nabla_{{\boldsymbol k}}   \xi_{ \chi, s } ({\boldsymbol k})
 = {\boldsymbol   v}_s ({\boldsymbol k} ) + {\boldsymbol  v}^{(m)}_{\chi, s} ({\boldsymbol k} ) \,,
\quad {\boldsymbol v}^{(m)}_{\chi, s} ({\boldsymbol k} )
= \nabla_{{\boldsymbol k}} \varepsilon_{\chi, s}^{(m)}   (\boldsymbol k) \,.
\end{align}
The effects of OMM show up via the modified energy appearing in the equilibrium Fermi-Dirac distribution,
\begin{align}
\label{eqdist}
	f_0 \big (\xi_{\chi, s} (\boldsymbol k) , \mu, T \big )
= \frac{1}
{ 1 + \exp [ \, 
\left(  \xi_{\chi, s} (\boldsymbol k)-\mu \right) /T  ]}\,,
\end{align}
where $T $ is the temperature.
While using $f_0$ in various equations, we will be suppressing its $\mu$- and $ T $-dependence for uncluttering of notations. Moreover, in what follows, we will restrict our calculations to the $ T = 0$ limit.

In our earlier work using RTA, appearing as Ref.~\cite{ips-spin1-ph}, we did not consider the quadratic corrections. Therefore, the flat-band there was ignored, considering the fact that a nonzero chemical potential (measured with respect to the nodal point) will not cut the band, leading to no contribution to conductivity. Of course the OMM would have given it a finite dispersion for a nonzero magnetic field --- but it would lead to open Fermi surfaces, which would have necessitated the introduction of artificial cut-offs in the momentum-integrals. Such an unphysical situation is remedied by introducing the quadratic-in-momentum corrections.

\subsection{Collinear electric and magnetic fields along the $z$-axis}

Let us write down the expressions for the OMM-induced energy- and velocity-corrections when $\boldsymbol B  =  B\, \boldsymbol{\hat{z}}$ is aligned along an external electric field, $ \boldsymbol E  =  E\, \boldsymbol{\hat{z}}$. For this case, we have
\begin{align}
\varepsilon_{\chi, s}^{(m)}   (\boldsymbol k)  = 
\frac{  \chi \, e\, B \, v_F \, \mathcal{G}_s \,  k_z } 
{2 \, k^2 } \text{ and }
{\boldsymbol  u}^{(m)}_{\chi, s} ({\boldsymbol k} ) 
=  \frac{ -\, \chi \, e\, B \, v_F \,  \mathcal{G}_s   } { k^4 }
 \left \lbrace k_z \, k_x, \, k_y \, k_z, \,  \frac{k_z^2 -k_x^2 - k_y^2 } {2} \right \rbrace .
\end{align}

The Hamilton's equations of motion for the chiral quasiparticles, subjected to collinear electric ($\boldsymbol{E}$) and magnetic ($\boldsymbol{B}$) fields, are given by \cite{mermin, sundaram99_wavepacket, li2023_planar, ips-kush-review, ips-rsw-ph, ips-shreya}
\begin{align}
\label{eqrkdot}
\dot {\boldsymbol r} & = \nabla_{\boldsymbol k} \, \xi_{\chi, s}  
- \dot{\boldsymbol k} \, \cross \, \boldsymbol \Omega_{\chi, s}  (\boldsymbol k)
 \text{ and }  
\dot{\boldsymbol k} = -\, e  \left( {\boldsymbol E}  
+ \dot{\boldsymbol r} \, \cross\, {\boldsymbol B} 
	\right ) \nn
\Rightarrow & \, \dot{\boldsymbol r}  = \mathcal{D}_{\chi, s} ({\boldsymbol k})
	\left[   \boldsymbol{w}_{\chi, s} ({\boldsymbol k}) +
	 e \, {\boldsymbol E}  \cross  
\boldsymbol \Omega_{\chi, s} ({\boldsymbol k})   + 
	 e   \left \lbrace \boldsymbol \Omega_{\chi, s} \cdot 
 \boldsymbol{w}_{\chi, s} ({\boldsymbol k}) 
 \right \rbrace  \boldsymbol B  \right] 
\nn & \text{ and } 
\dot{\boldsymbol k}  = -\, e \,\mathcal{D}_{\chi, s} ({\boldsymbol k})
  \left[   {\boldsymbol E} 
+   \boldsymbol{w}_{\chi, s} ({\boldsymbol k}) \cross  {\boldsymbol B} 
+  e  \left (  {\boldsymbol E}\cdot  {\boldsymbol B} \right )  
\boldsymbol \Omega_{\chi, s}  ({\boldsymbol k})\right],
\end{align}
where $-\, e$ is the quantum of charge for each quasiparticle. We note that a nonzero BC has modified the form of the Hamilton's equations of motion compared to the cases when the BC is zero (see, for example, the systems discussed in Refs.~\cite{ips-kush, ips_tilted_dirac}). Clearly, the $- \dot{\boldsymbol k} \, \cross \, \boldsymbol \Omega_{\chi, s} $ term in the expression for $\dot {\boldsymbol r}$ acts as an anomalous velocity, with the BC being an analogue of the magnetic field in the momentum space (comparing with the $ - e \, \dot{\boldsymbol r} \, \cross\, {\boldsymbol B} $ term appearing in
the expression for $\dot {\boldsymbol k}$). The kinetic equation, arising out of the fundamental Boltzmann formalism to compute transport, is expressed as
\begin{align}
\label{eqkin32}
& \left [
 \partial_t  
+ {\boldsymbol w}_{\chi, s} ({\boldsymbol k}) \cdot \nabla_{\boldsymbol r} 
-e \left \lbrace  \boldsymbol{E}
+  {\boldsymbol w}_{\chi, s} ({\boldsymbol k})  \times {\boldsymbol B} 
\right \rbrace  \cdot 
\nabla_{\boldsymbol k} \right ] f_{\chi, s} (\boldsymbol r, \boldsymbol k, t)
=  I_{\text{coll}} [f_{\chi, s} ({\boldsymbol k}) ]\,,
\end{align}
where $f_{\chi, s}  (\boldsymbol r, \boldsymbol k, t) $ represents the nonequilibrium quasipaticle-distribution function close to the semimetallic node of chirality $\chi$. The deviation, $ \delta f_{\chi, s} (\boldsymbol r, \boldsymbol k, t) \equiv 
f_{\chi, s}  (\boldsymbol r, \boldsymbol k, t) - f_0 (\xi_{\chi, s} (\boldsymbol k) )$, from the equilibrium distribution, $ f_0 (\xi_{\chi, s}(\boldsymbol k) )$, is caused by the probe field, $\boldsymbol E $, coupled with the nonquantising $\boldsymbol B$-field, whose magnitude is controlled by the magnitude of $\boldsymbol E$ (which itself is assumed to be small). Here, we restrict ourselves to time-independent and spatially-uniform external fields and, hence, $f_{\chi, s} $ must not depend on position and time, implying $ \delta f_{\chi, s}  (\boldsymbol r, \boldsymbol k, t) = \delta f_{\chi, s}  (\boldsymbol k)$. Expanding upto order $|\boldsymbol E|$ in smallness allows us to arrive at the linearised approximation for the resulting response-coefficients. On the right-hand side, $ I_{\text{coll}} [f_{\chi, s} ({\boldsymbol k}) ] $ symbolises the so-called \textit{collision integral}, which comprises the relevant scattering processes trying to relax $f_{\chi, s} ({\boldsymbol k})$ towards $f_0 (\xi_{\chi, s}({\boldsymbol k}))$.

For point-scattering mechanisms, the collision integral takes the form of
\begin{align}
I_{\text{coll}} [f_{\chi, s}({\boldsymbol k})]
 = \sum \limits_{\tilde \chi, \tilde s}
\int_{k'}
\mathcal{M}^{\chi, \tilde \chi}_{ s, \tilde s} (\boldsymbol k,\boldsymbol k^\prime)
\left[ f_{\tilde \chi , \tilde s} ( \boldsymbol k^\prime))  - f_{\chi, s} (\boldsymbol k) \right ] ,
\end{align}
where $\int_k \equiv \int d^{3}  \boldsymbol k  \, {\mathcal D}^{-1}_{\tilde \chi, \tilde s} ({\boldsymbol k} )
/ (2\,\pi)^3 $ denotes the three-dimensional integral in the momentum space, containing the modified phase-space factor due to the BC.
The scattering rate, parametrised by
\begin{align}
\mathcal{M}^{\chi, \tilde \chi}_{ s, \tilde s} (\boldsymbol k,\boldsymbol k^\prime) 
= \frac{2\, \pi \, \rho_{\rm imp} } {V} \,
\Big \vert \left \lbrace  \psi_{ \tilde \chi, \tilde s }({ \boldsymbol k^\prime }) \right \rbrace^\dagger 
\; {\mathcal V} ^{\chi, \tilde \chi}_{ s, \tilde s} (\boldsymbol k,\boldsymbol k^\prime) 
  \;  \psi_{ \chi, s }({ \boldsymbol k}) \Big \vert^2 \,
 \delta \Big( \xi_{\tilde \chi , \tilde s } (\boldsymbol k^\prime)
 - \xi_{  \chi , s } (\boldsymbol k ) \Big) \,,
\end{align}
is obtained by applying the Fermi's golden rule, where $ \rho_{\rm imp}$ represents the impurity-concentration (acting as the scattering centres), $V$ denotes the system's volume, and ${\mathcal V} ^{\chi, \tilde \chi}_{ s, \tilde s} (\boldsymbol k,\boldsymbol k^\prime) $ stands for the scattering-potential matrix (in the spinor space of the three-component wavevectors) convoluted to the momentum space. Here, we focus on elastic and nonmagnetic scatterings, for which $ {\mathcal V} ^{\chi, \tilde \chi}_{ s, \tilde s} (\boldsymbol k,\boldsymbol k^\prime) 
=  \mathbb{I}_{3 \times 3} \, {\mathcal V} ^{\chi, \tilde \chi}_{ s, \tilde s} $, which reduces to an identity matrix in the spinor space and does not have any momentum dependence.
Therefore, the scattering rate simplifies to
\begin{align}
\mathcal{M}^{\chi, \tilde \chi}_{ s, \tilde s} (\boldsymbol k,\boldsymbol k^\prime) 
= \frac{2\, \pi \, \rho_{\rm imp} 
\, |{\mathcal V} ^{\chi, \tilde \chi}_{ s, \tilde s} |^2} 
{V} \,
\Big \vert \left \lbrace  \psi_{ \tilde \chi, \tilde s }({ \boldsymbol k^\prime }) \right \rbrace^\dagger 
\; \psi_{ \chi, s }({ \boldsymbol k}) \Big \vert^2 \,
 \delta \Big( \xi_{\tilde \chi , \tilde s } (\boldsymbol k^\prime)
 - \xi_{  \chi , s } (\boldsymbol k ) \Big) \,.
\end{align}
Assuming an interchange-symmetry between the $\chi = \pm 1$ nodes and the values of the bands, we parametrise the associated scattering strengths as
\begin{align}
 \big |{\mathcal V} ^{1, -1}_{ s, \tilde s} \big |^2 =   \big |{\mathcal V} ^{ -1, 1}_{ s, \tilde s} \big|^2
 \equiv \frac{ 2\,\pi} {\rho_{\rm imp}}\, \beta^{\rm inter}_{s, \tilde s} \,,
 \quad  \big |{\mathcal V} ^{1, 1}_{ s, \tilde s} \big |^2  
 =  \big | {\mathcal V} ^{-1,- 1}_{ s, \tilde s} \big|^2
  \equiv  \frac{ 2\,\pi} {\rho_{\rm imp}} \, \beta^{\rm intra}_{s, \tilde s} \,,\quad
\beta^{\rm inter}_{1,0} =  \beta^{\rm inter}_{0,1} \,, \quad
\beta^{\rm intra}_{1,0} =  \beta^{\rm intra}_{0,1} \,.
\end{align}

Here, we need to solve the \textit{linearised Boltzmann equation}, captured by
\begin{align}
\label{eqkin5}
& - e\, {\mathcal D}_{\chi, s} ({\boldsymbol k})
 \left [
\left \lbrace {\boldsymbol{w}}_{\chi, s} ({\boldsymbol k})
+ e \,\Big(
{\boldsymbol \Omega}_{\chi, s} ({\boldsymbol k}) 
\cdot {\boldsymbol{w}}_{\chi, s}   ({\boldsymbol k})
 \Big )  \boldsymbol B \right \rbrace
\cdot {\boldsymbol E}  
\; \;	\frac{\partial  f_0 (\xi_{\chi, s}({\boldsymbol k})) }
 {\partial \xi_{\chi, s}({\boldsymbol k}) }
+  
\left \lbrace   {\boldsymbol{w}}_{\chi, s} ({\boldsymbol k})
\cross  {\boldsymbol B}  \right \rbrace
\cdot \nabla_{\boldsymbol k}
\, \delta f_{\chi, s} (\boldsymbol k) \right] 
 =  I_{\text{coll}} [f_{\chi, s} ({\boldsymbol k})] \,.
\end{align}
To solve the above equation, we parametrise the deviation as
\begin{align}
\label{eqansatz}
	\delta f_{\chi, s} (\boldsymbol {k}) =
-\,	 e\, 	\frac{\partial  f_0 (\xi_{\chi, s}) } {\partial \xi_{\chi, s} } 
	\,   {\boldsymbol E}  \cdot \bm{\Lambda}_{\chi, s} (\boldsymbol {k} )
=
-\,	 e\, 	\frac{\partial  f_0 (\xi_{\chi, s}({\boldsymbol k})) }
 {\partial \xi_{\chi, s} ({\boldsymbol k})} 
	\,   E \, {\Lambda}^z_{\chi, s} ( \boldsymbol {k} ) \,,
\end{align}
where $ \bm{\Lambda}_{\chi, s} (\boldsymbol {k} ) $ is the vectorial mean-free path. For our configuration, we get a nontrivial equation only for the $z$-component of $ \bm{\Lambda}_{\chi, s} ( \boldsymbol {k} ) $ as follows:
\begin{align}
\label{eqvec}
&  w^z_{\chi, s} ({\boldsymbol k})
+ e \, B \left [
{\boldsymbol \Omega}^\chi_{s} ({\boldsymbol k})
 \cdot {\boldsymbol{w}}_{\chi, s} ({\boldsymbol k})  \right ]
-\, 
e \, B \left [ {\boldsymbol{w}}_{\chi, s} ({\boldsymbol k}) \cross   
\boldsymbol{\hat z}  ({\boldsymbol k}) \right ] 
\cdot \nabla_{\boldsymbol k} {\Lambda}^z_{\chi, s} (\boldsymbol {k} )  
 = {\mathcal D}^{-1}_{\chi, s}  ({\boldsymbol k})
\sum \limits_{\tilde \chi, \tilde s}
\int_{k'}
\mathcal{M}^{\chi, \tilde \chi}_{ s, \tilde s} (\boldsymbol k,\boldsymbol k^\prime)
\left[ 
 {\Lambda}^z_{ \tilde \chi, \tilde s} (\boldsymbol {k}^\prime ) 
 -  {\Lambda}^z_{\chi, s} (\boldsymbol {k} )  
 \right ].
\end{align}
To solve the above equation, we take the self-consistent ansatz \cite{timm} that 
$ {\Lambda}^z_{\chi, s} \equiv {\Lambda}^z_{\chi, s} ( \mu, \theta)$ at an energy $\mu$, which only depends on the polar angle, $\theta$, and the chemical potential, $\mu$. This is because, for elastic scatterings, the integral over the full momentum space can be replaced by an integral over the Fermi surface at energy $\xi_{\chi , s} (\boldsymbol k) = \mu $, and the dependence on the azimuthal angle $\phi$ should not be there because of the rotational symmetry of the entire system in the $k_x k_y$-plane. 
Consequently, the momentum-space integrals reduce to the respective Fermi surfaces at energy $\xi_{\chi , s} 
(k_F^{\chi, s} , \theta )= \mu $ with $T$ set to zero, denoted by the set of Fermi momenta, $ \lbrace k_F^{\chi, s} (\theta) \rbrace $. The self-consistency can be easily checked from the fact that $\left [ {\boldsymbol \Omega}^\chi_{s} (\boldsymbol k) \cdot
 {\boldsymbol{w}}_{\chi, s} (\boldsymbol k)  \right ] $ is $\phi$-independent and $ \left [  {\boldsymbol{w}}_{\chi, s} 
 (\boldsymbol k)  \cross   \boldsymbol{\hat z}  \right ] 
\cdot \nabla_{\boldsymbol k} {\Lambda}^z_{\chi, s} (\mu, \theta ) $ evaluates to zero.

Going by the above understanding, for our set-up involving $ {\boldsymbol E}  \parallel {\boldsymbol B} $, 
although the spinor overlaps contain the azimuthal angles,
$\phi$ and $\phi^\prime $, they will drop out on performing the azimuthal-angle integrations. Hence, using Eq.~\eqref{eqev},we can effectively use the overlap-function defined as 
\begin{align}
\label{eqoverlap}
{\mathcal T }^{\chi, \tilde \chi}_{ s, \tilde s} (\theta, \theta^\prime)
& = \left [
\sin ^4\bigg(\frac{\theta }{2}\bigg) \, \sin ^4\bigg (\frac{\theta '}{2}\bigg )
+\frac{1}{4} \sin ^2 \theta  \, \sin ^2 \theta'
+ \cos ^4\bigg(\frac{\theta }{2}\bigg ) \cos^4\bigg (\frac{\theta '}{2}\bigg )
\right ] \delta_{s,1} \, \delta_{\tilde s, 1}
 \nn & \quad 
+ \left [
\frac{ \sin ^2 \theta \, \sin ^2 \theta ' }{2} 
+\cos ^2 \theta \, \cos ^2 \theta '
\right ] \delta_{s,0} \, \delta_{\tilde s, 0}
\nn & \quad + \left [ 
\left\lbrace \sin ^4\bigg (\frac{\theta }{2}\bigg )
+ \cos ^4\bigg( \frac{\theta }{2}\bigg )
\right \rbrace \sin ^2 \theta '
+\sin ^2 \theta \, \cos ^2\theta '  \right ]   \delta_{s,1} \, \delta_{\tilde s, 0} 
\nn & \quad + \left [ 
 \sin ^2 \theta 
\left\lbrace \sin ^4\bigg (\frac{\theta^\prime }{2}\bigg )
+ \cos ^4 \bigg( \frac{\theta ^\prime}{2}\bigg )
\right \rbrace
+\cos ^2\theta \, \sin ^2 \theta^\prime  \right ]  
\delta_{s,0} \, \delta_{\tilde s, 1} \,,
\end{align}
while evaluating the integrals. Using the $\phi$-integrated forms, Eq.~\eqref{eqvec} reduces to
\begin{align}
\label{eq_lambda_mu}
h_{\chi, s} (\mu, \theta) = 
 \sum_{\tilde \chi , \tilde s} V  
\int_{k^\prime }\, \mathcal{M}^{\chi, \tilde \chi}_{ s, \tilde s} 
(\boldsymbol k,\boldsymbol k^\prime)
\, {\Lambda}^z_{ \tilde \chi, \tilde s} ( \mu, \theta^\prime ) 
 - \frac{ {\Lambda}^z_{\chi, s}  (\mu, \theta)} 
{\tau_{\chi, s}(\mu, \theta)}  \,,
\end{align}
where
\begin{align}
\tau^{-1}_{\chi, s}(\mu, \theta)
= \sum_{\tilde \chi , \tilde s} V \int_{k^\prime }
\mathcal{M}^{\chi, \tilde \chi}_{ s, \tilde s} (\boldsymbol k,\boldsymbol k^\prime) \,, \quad
h_{\chi, s} (\mu, \theta) = {\mathcal D}_{\chi, s} ({\boldsymbol k})
 \left[ w^z_{\chi, s} ({\boldsymbol k})
+ e \, B \left \lbrace
{\boldsymbol \Omega}_{\chi, s} ({\boldsymbol k}) \cdot {\boldsymbol{w}}_{\chi, s} ({\boldsymbol k})
  \right \rbrace \right].	
\end{align}
Since the integrals reduce to the respective Fermi surfaces at energy $ \mu $ with $T$ set to zero, the above expression further simplifies to
\begin{align}
\label{eqlambdamu}
& h_{\chi, s}(\mu, \theta) + \frac{ {\Lambda}^z_{\chi, s}  (\mu, \theta)} 
{\tau_{\chi, s}(\mu, \theta)} =
\sum_{\tilde \chi , \tilde s}  
\frac{ \rho_{\rm imp} 
\, |{\mathcal V} ^{\chi, \tilde \chi}_{ s, \tilde s} |^2 }
{ 4\, \pi }
\int d\theta^\prime \, \frac{\sin \theta^\prime \left (k^\prime \right )^3 
\, {\mathcal D}^{-1}_{\tilde \chi, \tilde s} ({\boldsymbol k}^\prime)}
{  |\boldsymbol k^\prime \cdot {\boldsymbol{w}}_{\chi, s} (\boldsymbol k^\prime)  | }
\, {\mathcal T }^{\chi, \tilde \chi}_{ s, \tilde s} (\theta, \theta^\prime)
\, {\Lambda}^z_{ \tilde \chi, \tilde s} (\mu, \theta^\prime ) 
\Big \vert_{ k^\prime = k_F^{ \tilde \chi, \tilde s} } \,.
\end{align}
While the factor $ \left (k^\prime \right )^2  \sin \theta^\prime $ arises as the Jacobian for switching to the spherical polar coordinates, the part $$  \big |\boldsymbol {\hat k^\prime} \cdot 
 \nabla_{\boldsymbol k^\prime} \xi_{\chi, s}({\boldsymbol k^\prime}) \big |^{-1}  
 = k^\prime / |\boldsymbol k^\prime \cdot {\boldsymbol{w}}_{\chi, s} (\boldsymbol k^\prime)  |$$
arises from converting $ \delta \Big( \xi_{\tilde \chi , \tilde s } (\boldsymbol k^\prime) -\mu \Big) $ to
$\delta (k^\prime - k_F^{ \tilde \chi, \tilde s})$.

\begin{figure*}[]
\subfigure[]{\includegraphics[width= 0.3 \textwidth]{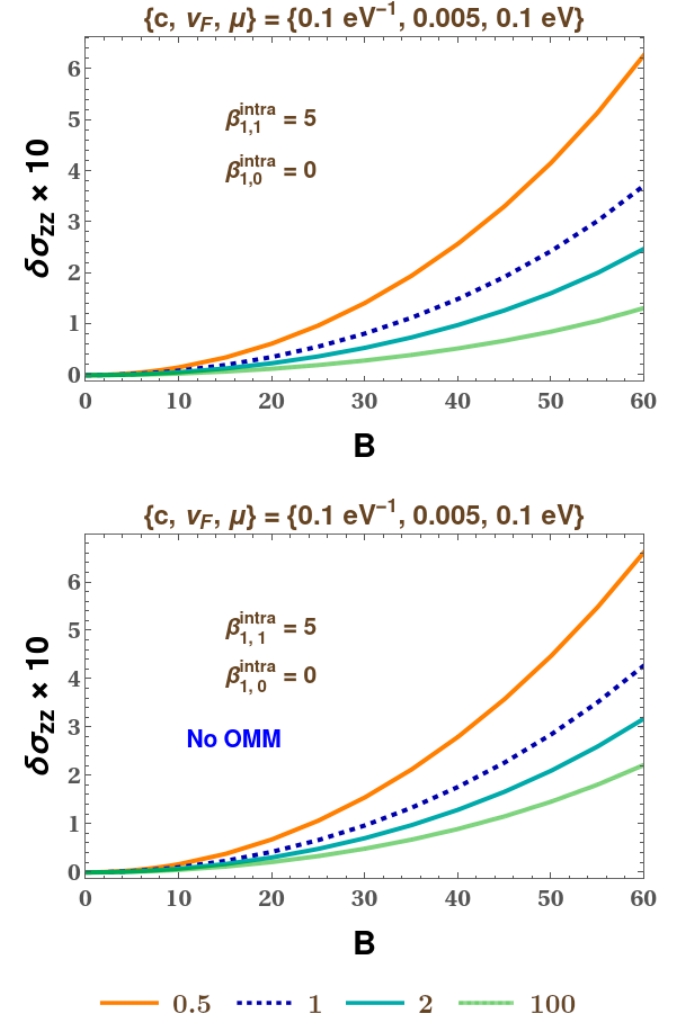}} \quad \vrule
\subfigure[]{\includegraphics[width= 0.3 \textwidth]{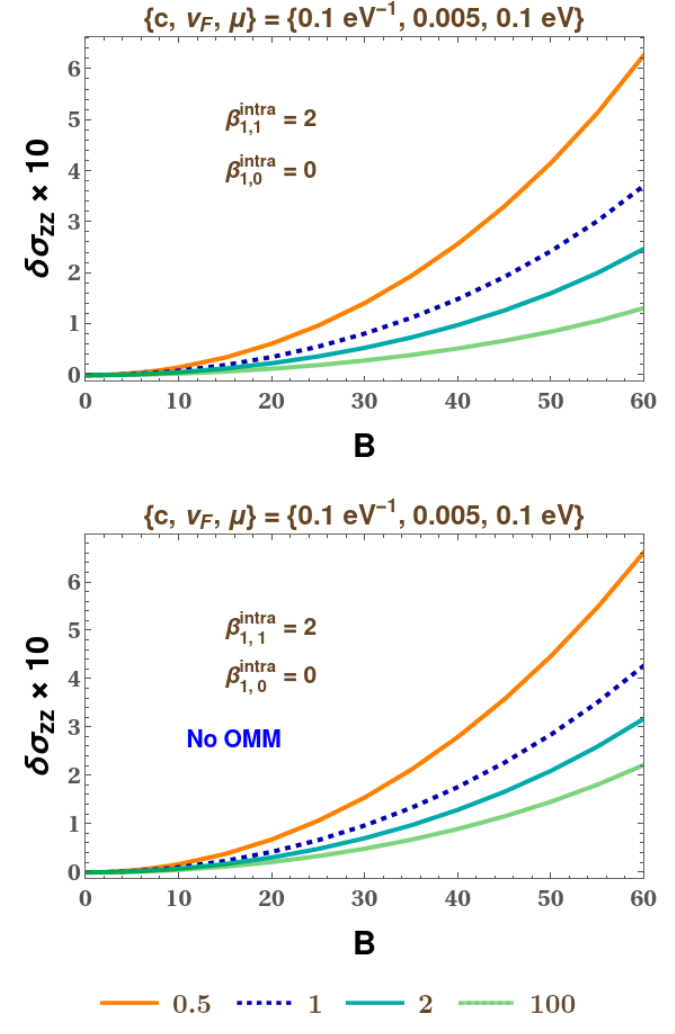}}\quad \vrule
\subfigure[]{\includegraphics[width= 0.31 \textwidth]{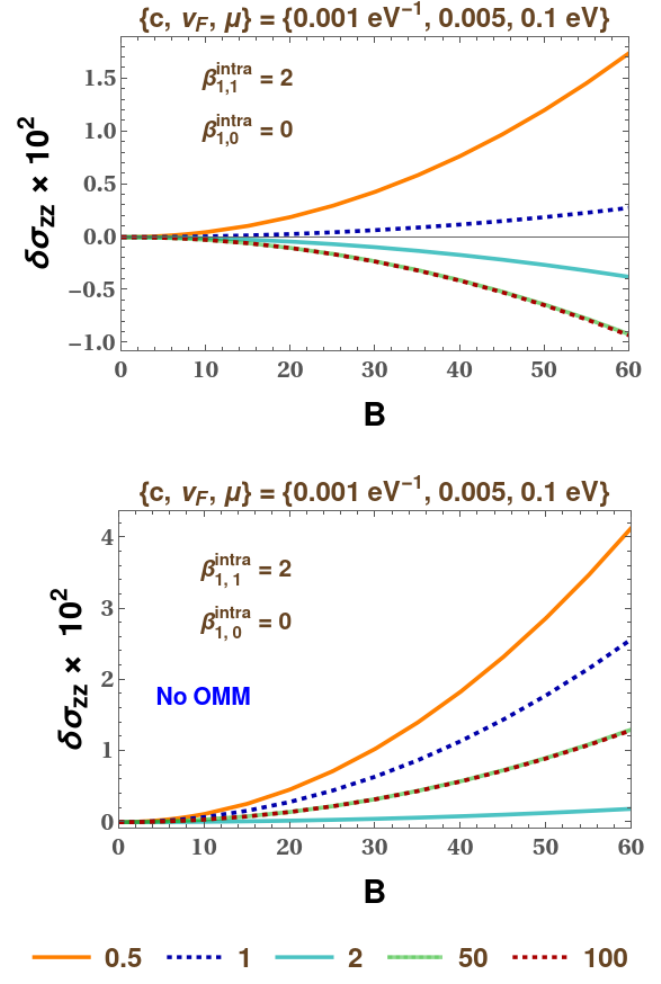}} 
\caption{\label{figsep1}$\delta \sigma_{zz}$ from the $\{s, \tilde s\} = \{1,1\}$ bands with no interaction with the $\{s, \tilde s\} = \{0,0\}$ bands: While the top panel shows the variation of the full conductivity with $B$ (in eV$^2$) when OMM is taken into account appropriately, the bottom panel represents conductivity versus $B$ when OMM is not considered. The plot-legends indicate the values of the ratio $\beta^{\rm inter}_{1,1} / \beta^{\rm intra}_{1,1}$. The three subfigures represent three distinct sets of parameter values, as indicated in the labels.}
\end{figure*}

Noting the form of the wavevectors and their overlaps [cf. Eqs.~\eqref{eqvec} and \eqref{eqoverlap}],
we now make the following ansatz:
\begin{align}
 {\Lambda}^z_{\chi, 1} (\mu,\theta ) & = 
\tau_{\chi, 1}(\mu,\theta)
 \left [ - h_{\chi, 1} (\mu, \theta) 
 + a_{\chi, 1} \, \cos^4\bigg(\frac{\theta }{2}\bigg)
+ b_{\chi, 1} \,\sin^4\bigg(\frac{\theta }{2}\bigg) 
+ c_{\chi, 1} \, \sin^2 \theta  \right ],\nn
{\Lambda}^z_{\chi, 0} (\mu,\theta ) & = 
 \tau_{\chi, 0} (\mu,\theta)
 \left [ -h_{\chi, 0} (\mu,\theta) 
 + c_{\chi, 0} \,\sin^2 \theta
+ d_{\chi, 0} \, \cos^2 \theta  \right ],
\end{align}
corresponding to the two positive-energy bands at each node. See also Appendix~\ref{appansatz} for understanding the logic behind this choice. This leaves us with the problem of solving for the 10 unknown
coefficients, $\lbrace a_{\chi, 1}  , \, b_{\chi, 1} , \,  c_{\chi, 1} ,
\,c_{\chi, 0}, \,d_{\chi, 0} \rbrace $, which must be real numbers. Plugging in the ansatz in Eq.~\eqref{eqlambdamu} furnishes
ten linear equations, which can be written as a matrix equation of the form
\begin{align}
\label{eqmatrix}
\mathcal A \, \mathcal C = \mathcal H \,, \text{ where }
\mathcal C =\begin{bmatrix}
a_{1, 1} & b_{1, 1}  & c_{1, 1} & c_{1, 0} & d_{1, 0} &
a_{-1, 1} & b_{-1, 1}  & c_{-1, 1} & c_{-1, 0} & d_{-1, 0} 
\end{bmatrix}^{\rm T}\,.
\end{align}
The explicit forms of the square and column matrices, $ \mathcal A $ and $ \mathcal H $, are detailed in Appendix~\ref{appmat}.
Additionally, the electron-number conservation furnishes the constraint of
\begin{align}
\label{eqcon}
\sum \limits_{\chi, s}	\int_k  \delta f_{\chi, s} (\boldsymbol {k}) = 0\,.
\end{align}

Inserting the solutions, we obtain the charge-current density along the $z$-direction as
\begin{align}
\label{def_cur}
	J_z^{\rm tot} =- \,\frac{e}{ V }\sum_{\chi,s}
\int_k	(\dot{\boldsymbol {r}} \cdot \boldsymbol{\hat z})
\;  \delta f_{\chi, s} (\boldsymbol k)\,,
\end{align}
leading to
\begin{align}
\sigma_{zz}^{\rm tot} =  -\,\frac{e^2 } { V } 
\sum_{\chi,s}
\int \frac{d^3 {\boldsymbol k}} {(2\, \pi)^3} 
	\left[   w^z_{\chi, s} (\boldsymbol k)
+ e \, B  \left \lbrace \boldsymbol \Omega_{\chi, s} (\boldsymbol k)
\cdot 	  \boldsymbol{w}_{\chi, s} (\boldsymbol k) \right  \rbrace   \right] 
\delta \big (\xi_{\chi, s} (\boldsymbol k)-\mu \big) \, {\Lambda}^z_{\chi, s} (\mu, \theta )	\,. 
\end{align}

\begin{figure*}[]
\subfigure[]{\includegraphics[width= 0.3 \textwidth]{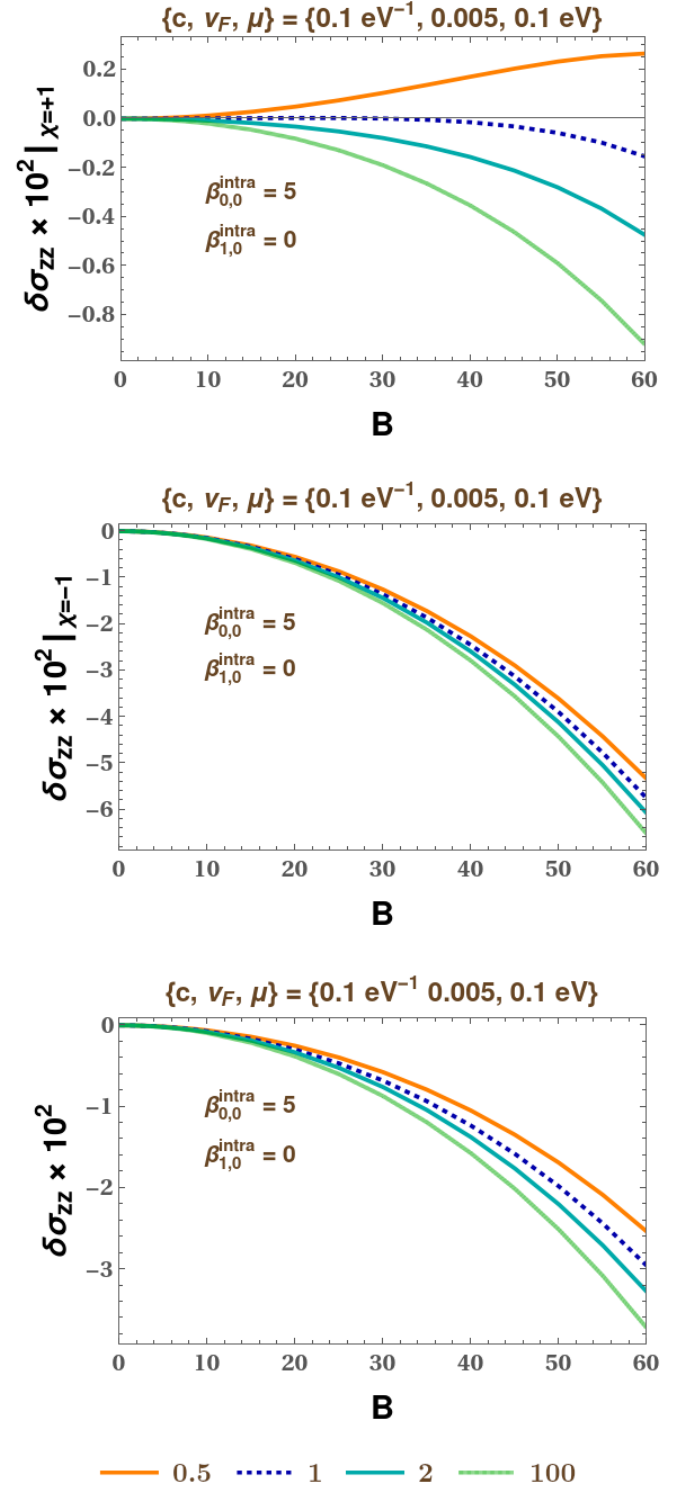}} \quad \vrule
\subfigure[]{\includegraphics[width= 0.3 \textwidth]{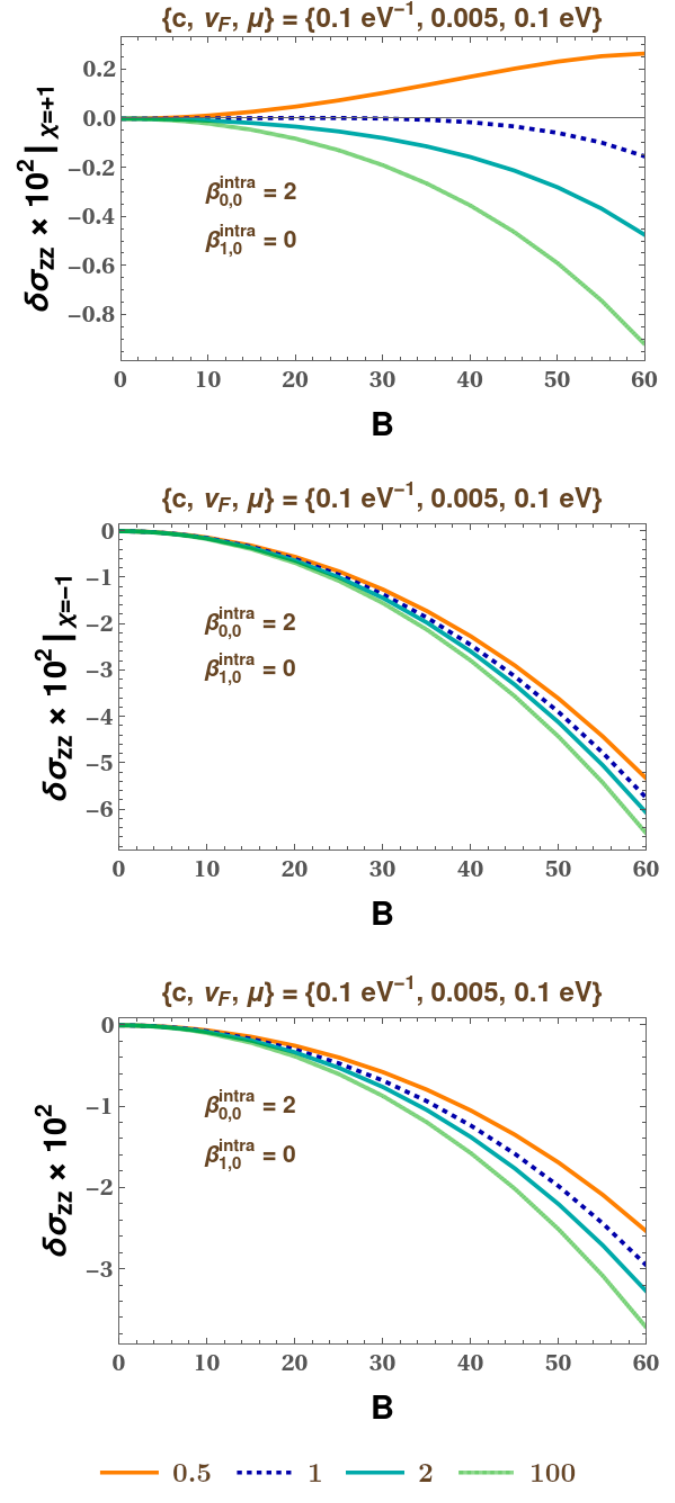}} \quad \vrule
\subfigure[]{\includegraphics[width= 0.305 \textwidth]{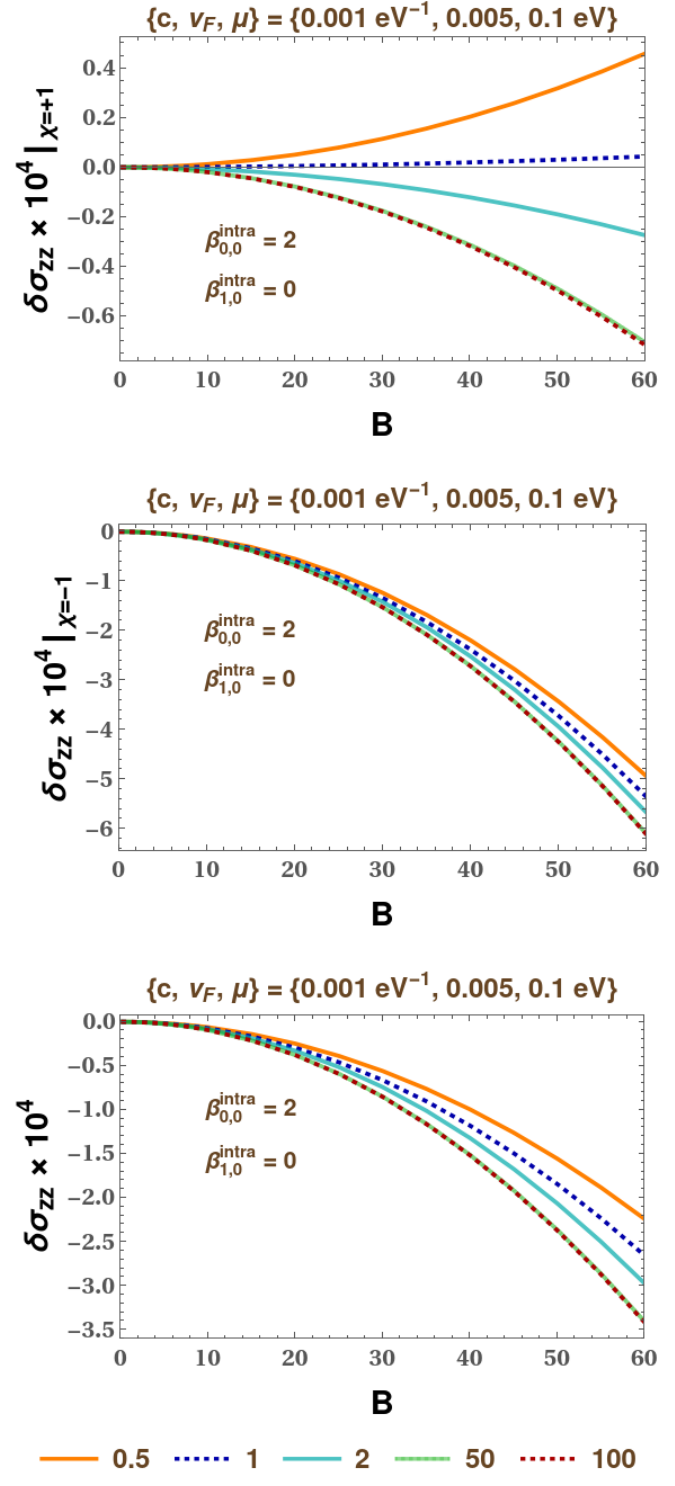}} 
\caption{\label{figsep2}$\delta \sigma_{zz}$ from the $\{s, \tilde s\} = \{0,0\}$ bands with no interaction with the $\{s, \tilde s\} = \{1,1\}$ bands: The curves demonstrate the variation of conductivity with $B$ (in eV$^2$). Here, the $B$-dependence is entirely caused by the OMM, since BC is zero for these bands. The top, middle, and bottom panels represent the results for the $\chi=+1$ node, $\chi =-1$ node, and sum over both the nodes, respectively. The plot-legends indicate the values of the ratio $\beta^{\rm inter}_{0,0} / \beta^{\rm intra}_{0,0}$. The three subfigures represent three distinct sets of parameter values, as indicated in the labels.}
\end{figure*}

\begin{figure*}[]
\subfigure[]{\includegraphics[width= 0.3 \textwidth]{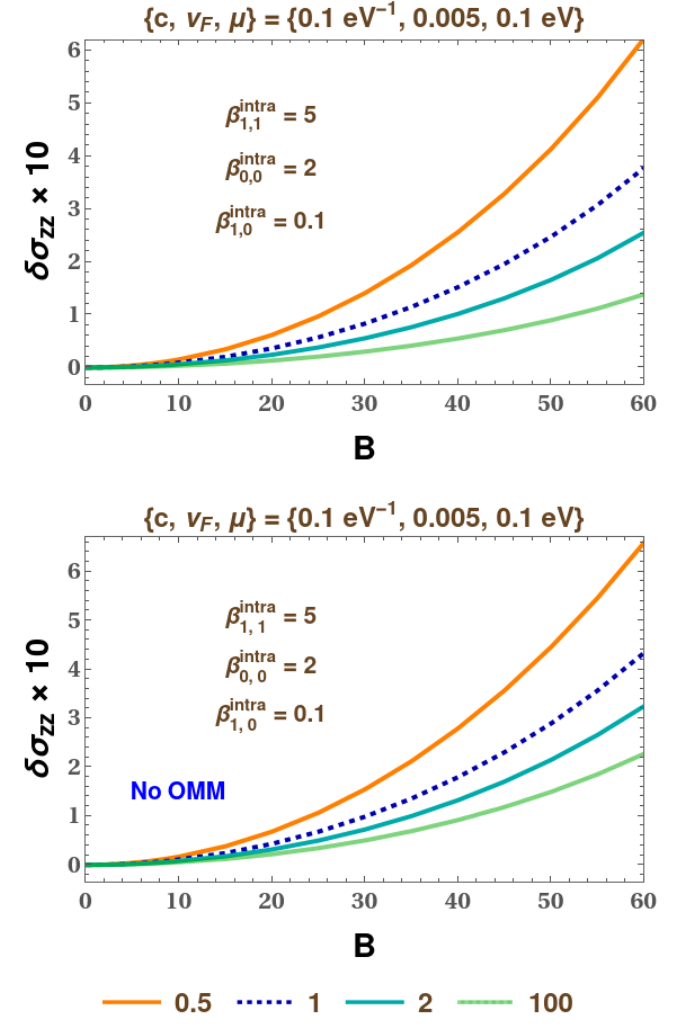}} \quad \vrule
\subfigure[]{\includegraphics[width= 0.3 \textwidth]{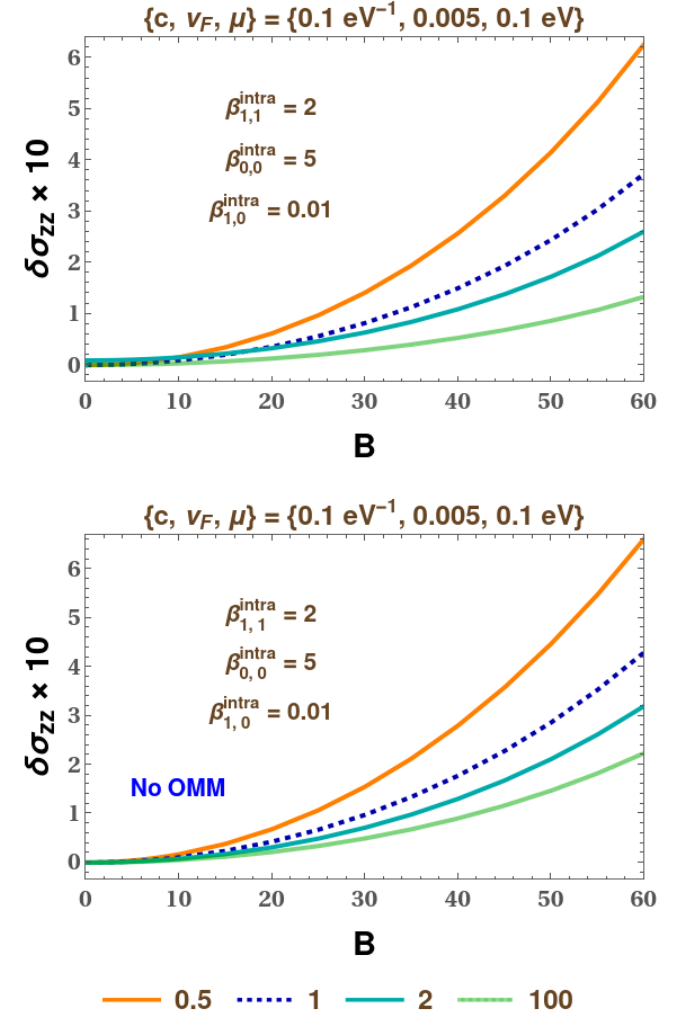}}\quad \vrule
\subfigure[]{\includegraphics[width= 0.3 \textwidth]{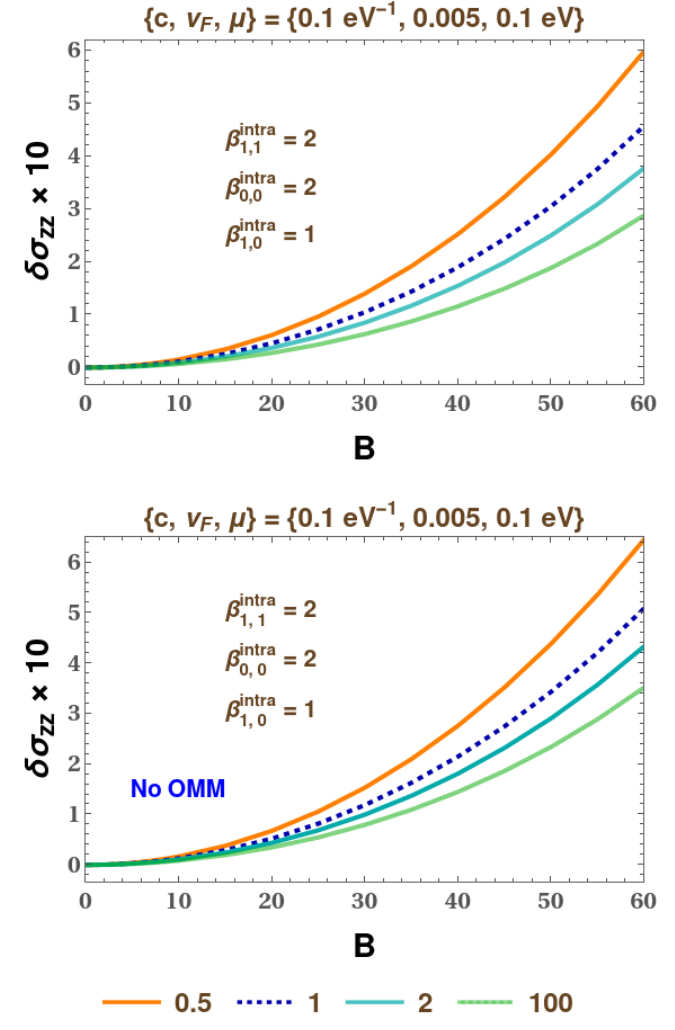}}  
\caption{\label{figall1}$\delta \sigma_{zz}$ from the $\{s, \tilde s\} = \{1,1\}$ bands interacting mutually and with the $\{s, \tilde s\} = \{0,0\}$ bands: While the top panel shows the variation of the full conductivity against $B$ (in eV$^2$) when OMM is taken into account appropriately, the bottom panel represents conductivity versus $B$ when OMM is omitted. The plot-legends indicate the values of the ratios $\lbrace \beta^{\rm inter}_{s,\tilde s} / \beta^{\rm intra}_{s,\tilde s}
\rbrace $. The three subfigures represent three distinct sets of parameter values, as indicated in the labels.}
\end{figure*}

\begin{figure*}[]
\subfigure[]{\includegraphics[width= 0.3 \textwidth]{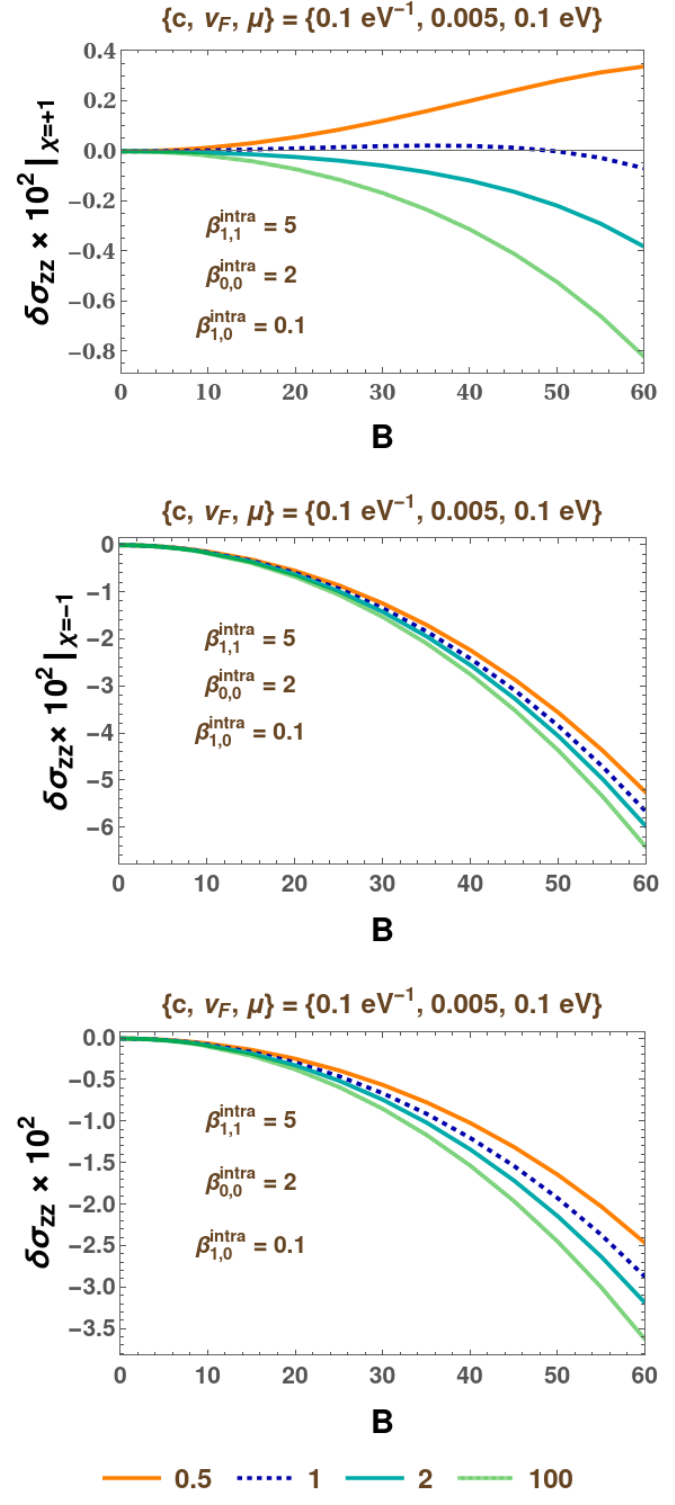}} \quad \vrule
\subfigure[]{\includegraphics[width= 0.3 \textwidth]{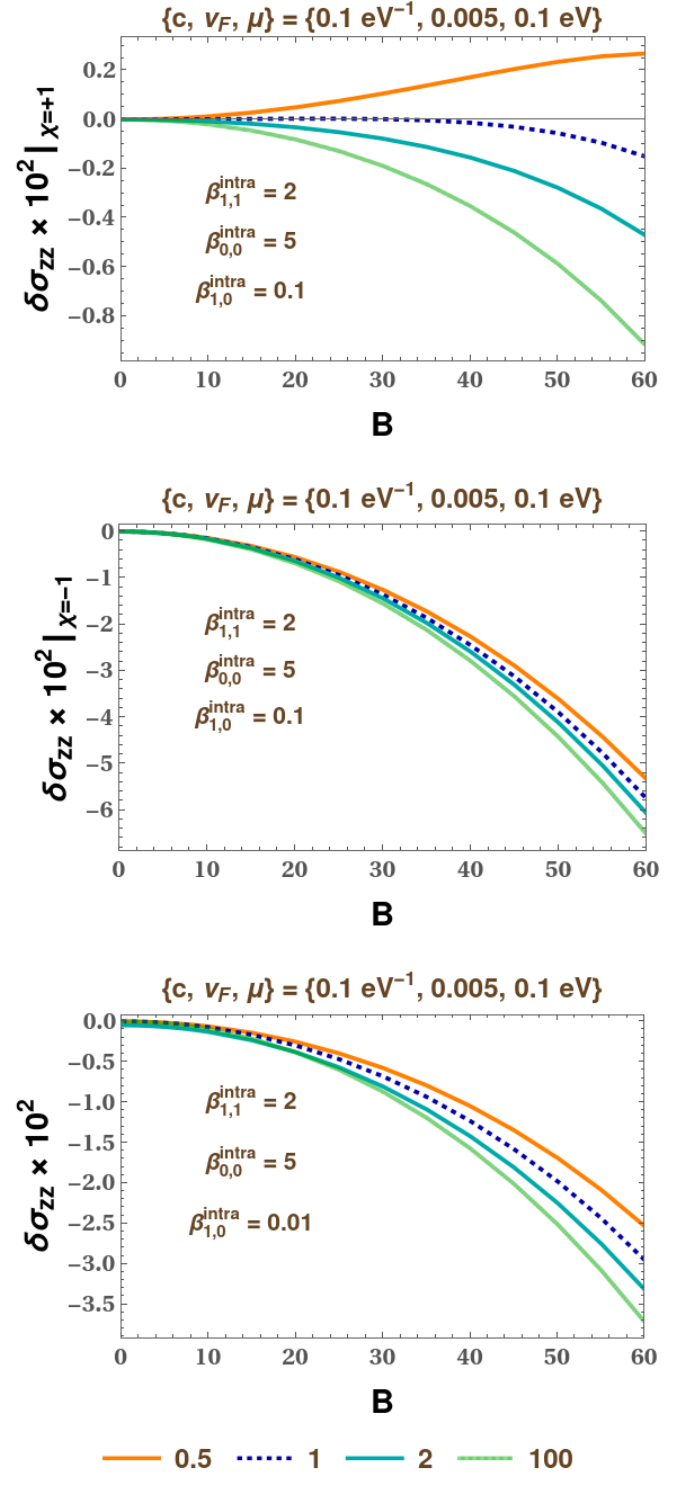}} \quad \vrule
\subfigure[]{\includegraphics[width= 0.3 \textwidth]{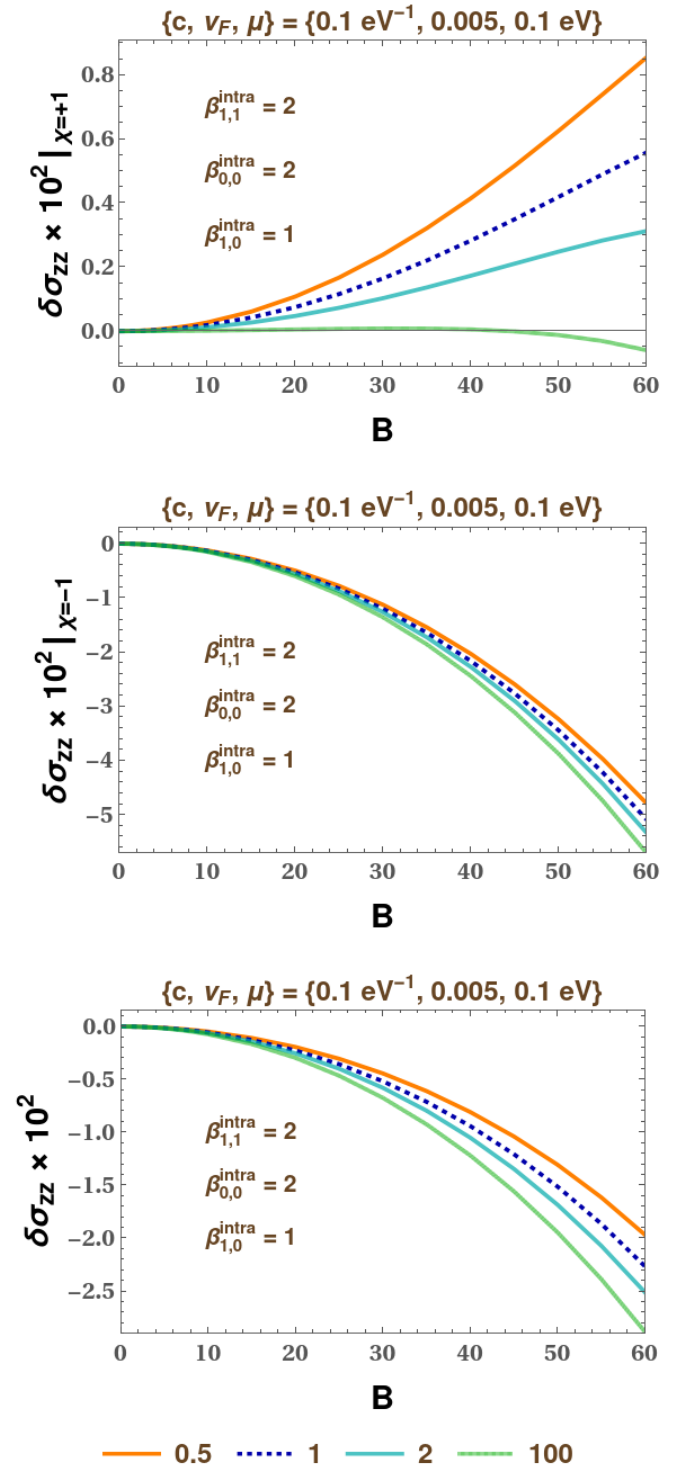}} 
\caption{\label{figall2}$\delta \sigma_{zz}$ from the $\{s, \tilde s\} = \{0,0\}$ bands interacting mutually and with the $\{s, \tilde s\} = \{1,1\}$ bands: The curves demonstrate the variation of conductivity  with $B$ (in eV$^2$). The top, middle, and bottom panels represent the results for the $\chi=+1$ node, $\chi =-1$ node, and sum over both the nodes, respectively. The plot-legends indicate the values of the ratios $\lbrace \beta^{\rm inter}_{s,\tilde s} / \beta^{\rm intra}_{s,\tilde s} \rbrace $. The three subfigures represent three distinct sets of parameter values, as indicated in the labels.}
\end{figure*}

\subsection{Results and discussions} 
\label{secres}

\begin{figure*}[]
\subfigure[]{\includegraphics[width= 0.6 \textwidth]{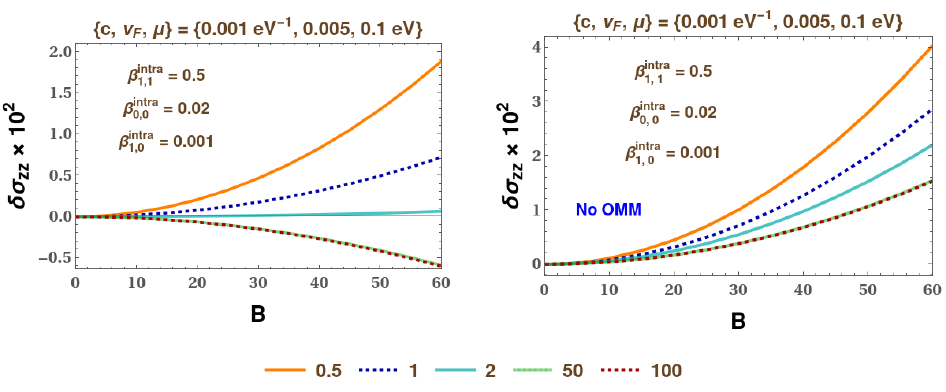}} 
\subfigure[]{\includegraphics[width= 0.9 \textwidth]{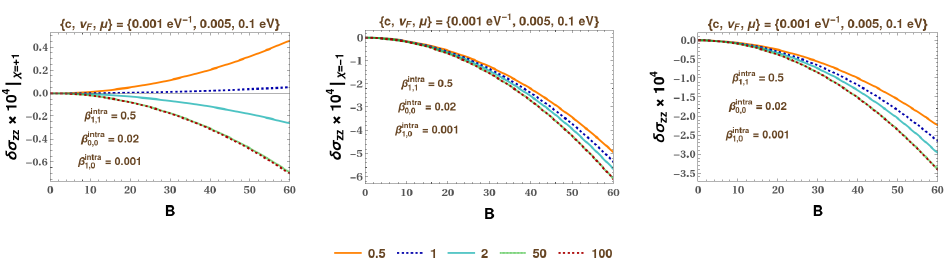}} 
\caption{\label{figall3}(a) Net $\delta \sigma_{zz}$ from the $\{s, \tilde s\} = \{1,1\}$ bands, when interacting mutually and with the $\{s, \tilde s\} = \{0,0\}$ bands, as $B$ (in eV$^2$) is cranked up. (b) Node-dependent and net values of $\delta \sigma_{zz}$ for the $\{s, \tilde s\} = \{0,0\}$ bands, as $B$ is varied, when interacting mutually and with the $\{s, \tilde s\} = \{ 1,1 \}$ bands. The plot-legends indicate the values of the ratios $\lbrace \beta^{\rm inter}_{s,\tilde s} / \beta^{\rm intra}_{s,\tilde s} \rbrace $.}
\end{figure*}

We now discuss the numerical results obtained by solving Eq.~\eqref{eqlambdamu}. For each case, we demonstrate the behaviour of $$\delta \sigma_{zz}  \equiv \sigma_{zz} (B) / \sigma_{zz}(B=0)  -1 $$ for some representative parameter values, which will help visualising the interplay between the values of $c$ and $ \lbrace \beta^{\rm intra}_{s, \tilde s}, \,\beta^{\rm inter}_{s, \tilde s} \rbrace $. Depending on the context, $\sigma_{zz} (B)$ will represent the contribution from one particular type of band (i.e., with a particular value of $S_z$ quantum number) from one individual node or both the nodes. For each of $s $ and $\tilde s$ taking the value of zero, the entire $B$-dependent conductivity is caused by the OMM, because the BC is zero for these bands. However, for the linear bands with small quadratic corrections, both the BC and the OMM contribute. Therefore, for these bands, in order to determine what the error would amount to if the OMM is not considered, we provide comparison-curves representing the cases when OMM is omitted. For all our plots, we have set $ v_F = 0.005 $ and $ \mu = 0.1 $ eV. Note that $c$ is in eV$^{-1}$. 

First, let us consider the case when the scattering processes obey the conditions $ \beta^{\rm intra}_{s, \tilde s}  \propto \delta_{s, \tilde s}$ and $ \beta^{\rm inter}_{s, \tilde s} \propto \delta_{s, \tilde s}$, i.e., the quasiparticles carrying pseudospin-projection values of 1 and 0 do not talk to each other. Consequently, $ \mathcal A $ decomposes into the direct sum of the two matrices,
$ \mathcal{A}_1 $ and $ \mathcal{A}_0 $, where the former (latter) is a $6$-dimensional ($4$-dimensional) square matrix. In every case, we find that $ \mathcal{A}_1 $ has rank $5$ (instead of $6$) and, hence, needs to be supplemented by the conservation equation, $\sum _{\chi}	\int_k  \delta f_{\chi, 1} (\boldsymbol {k}) = 0$, to determine all the unknown coefficients. Similarly, $ \mathcal{A}_0 $ has rank $3$ (instead of $4$) and the the conservation equation, $\sum _{\chi}	\int_k  \delta f_{\chi, 0} (\boldsymbol {k}) = 0$, must be used to determine all the unknown coefficients. Figs.~\ref{figsep1} and \ref{figsep2} capture such a scenario for the bands $1$ and $0$, respectively, each curve showing a $\propto B^2$-dependence at leading order. For the linear bands with small quadratic corrections, in each case, the BC-only and OMM-induced parts have opposite signs. In fact, when $ \beta^{\rm inter}_{1,1} / \beta^{\rm intra}_{1,1} > 1$ and is bigger than a threshold, the OMM acquires a large magnitude, thus managing to flip the sign of $\delta \sigma_{zz}$, which also crucially depends on the values of $c$ [cf. Fig.~\ref{figsep1} (c)]. For the conductivity arising from the $s = \tilde s = 0$ bands, sourced only by the OMM, $\delta \sigma_{zz}$ (representing the sum over both the nodes) is always negative. We also look into the behaviour a bit deeper by disentangling the contributions for the $\chi =+1$ and $\chi =-1$ nodes separately, originating from the  $s = \tilde s = 0$ bands. In general, we find that they are asymmetric/unequal for a nonzero $B$, with the magnitude of $\delta \sigma_{zz} \vert_{\chi= -1}$ dominating over that of $\delta \sigma_{zz} \vert_{\chi= + 1}$. While $\delta \sigma_{zz} \vert_{\chi= -1}$ always remain negative, for optimally low values of the internode-scattering amplitude, we find that $\delta \sigma_{zz} \vert_{\chi= + 1}$ is positive. The latter behaviour can be nicely fitted with a curve of the form $ a \, B^2 - b \, B^4$, where $a>0$ and $b>0 $. However, the total, $ \sum \chi \,\delta \sigma_{zz} \vert_{\chi} $ is always negative, as reported above.
 
Next, we consider the case when quasiparticles of all the participating conduction bands can scatter with one another. Figs.~\ref{figall1},  \ref{figall2}, and \ref{figall3} are the illustrations representative for this scenario. In particular, Fig.~\ref{figall3} captures the behaviour for a low ratio of $c/v_F$, corresponding to the parameters used in Figs.~\ref{figsep1}(c) and Figs.~\ref{figsep2}(c). On explicit computations, the $10$-dimensional square matrix $ \mathcal{A}$ is found to have rank $9$. Therefore, Eq.~\eqref{eqcon} must be used to determine all the unknown coefficients. Here too we find that, for the linear bands with small quadratic corrections, the OMM-contribution has a sign opposite to that of the BC-only-induced parts. For the $s = \tilde s = 0$ bands, the asymmetric $\chi$-dependent characteristics persist, such that $\delta \sigma_{zz} \vert_{\chi= -1}$ always remain negative and, for low-enough values of internode-scattering amplitude, we find that $\delta \sigma_{zz} \vert_{\chi= + 1}$ has positive values. Furthermore, a high value of $\beta^{\rm intra}_{1,0}  $ pushes $\delta \sigma_{zz} \vert_{\chi= + 1}$ towards positive values as well [cf. Fig.~\ref{figall2}(c)]. Summing over the two nodes yields an overall negative value of $\delta \sigma_{zz}$ for these bands. Hence, our trial datasets reveal an incredible amount of tunability in obtaining a desired nature of
the conductivity.

Let us now dissect the origins of various contributions to the net magnetoconductance, trying to pin-point its sign upon including the OMM:
\begin{itemize}
\item Case I --- The features of $\delta \sigma_{zz}$ must be the same as in the WSM case when we consider only the $s = \tilde s = 1$ bands, exemplified by Fig.~\ref{figsep1}(c). Let us label the BC-only, OMM-only, and the concurrent part of $ \delta \sigma_{zz} $ as $ \delta \sigma_{zz}^{BC}$, $\delta \sigma_{zz}^{OM}$, and $ \delta \sigma_{zz}^{conc}$, respectively. The last one implies the part when both the BC and OMM contribute together. Indeed, we find that the sign change of $\delta \sigma_{zz}$ is not the result of a simple competition between $ \delta \sigma_{zz}^{BC}$ and $ \delta \sigma_{zz}^{OM}$, as each of these two mechanisms individually produces a positive value. Only $ \delta \sigma_{zz}^{conc}$ comes with a negative sign. This is because  of the following: First we observe that $\eta_{\chi, bc} \equiv \text{sgn}[\boldsymbol \Omega_{\chi,1}  \cdot \boldsymbol B ]  $ and $\eta_{\chi, m} \equiv \text{sgn}[\varepsilon_{\chi, 1}^{(m)} ] $ have opposite signs. Since the response always goes as $B^2$, the surviving individual contributions appear as $\eta_{\chi, bc}\, \eta_{\tilde \chi, bc} $ and $\eta _{\chi ,m} \, \eta_{\tilde{\chi}, m}$, respectively, each of which has a sign opposite to that of $ \eta_{\chi, bc} \, \eta_{\tilde \chi , m} $ appearing in $ \delta \sigma_{zz}^{conc}$. As we crank up $ \beta^{\rm inter}_{1,1} / \beta^{\rm intra}_{1,1} $, $ |  \sigma_{zz}^{conc} |$ increases and $  \delta\sigma_{zz}^{BC} + \delta\sigma_{zz}^{OM}$ decreases in magnitude, with the former overtaking the latter eventually. For example, we find the following dataset for $B = 60 $ eV$^2$:
\\$ \delta \sigma_{zz} = 0.0174257 $, $ \delta \sigma_{zz}^{BC}  = 0.0413964 $,  
$\delta \sigma_{zz}^{OM}  =  0.0377579 $, and
$\delta \sigma_{zz}^{conc} = -0.0617285$ for $\beta^{\rm inter}_{1,1} / \beta^{\rm intra}_{1,1} = 0.5$;
\\$ \delta \sigma_{zz} = 0.00277429 $, $\delta \sigma_{zz}^{BC} = 0.0257312 $,  
$\delta \sigma_{zz}^{BC}  = 0.0204378 $, and
$\delta\sigma_{zz}^{conc}  = -0.0433948 $ for $\beta^{\rm inter}_{1,1} / \beta^{\rm intra}_{1,1} = 1 $;
\\$ \delta \sigma_{zz} = -0.00376071 $, $\delta \sigma_{zz}^{BC} = 0.0186851 $,  
$ \delta \sigma_{zz}^{BC}  = 0.0122026 $, and
$ \delta \sigma_{zz}^{conc} = - 0.0346485 $ for $\beta^{\rm inter}_{1,1} / \beta^{\rm intra}_{1,1} = 2$.
\item  Case II --- For the case when we consider only the subsystem of $s = \tilde s = 0$ bands, exemplified by the lowermost row of Fig.~\ref{figsep2}, we note that we have no BC contribution whatsoever. The overall $\delta \sigma_{zz} $ is always negative irrespective of the value of $\beta^{\rm inter}_{0,0} / \beta^{\rm intra}_{0,0}$. We also note that the magnitude of $\delta \sigma_{zz}$ is roughly of the order of 10 or 100 times smaller than that for Case I.
\item Case III --- When the interplay of all the bands are included, the qualitative features of Case I and Case II continue to persist when we look at the contributions coming from the two types of bands.
By observing that the magnitude of $\delta \sigma_{zz}$ is order(s) of magnitude higher for the $s = \tilde s = 1$ bands compared to the $s = \tilde s = 0$ bands, the net conductivity coming from all the bands (which is what will be measured in an experiment) will be dominated by the former.
\end{itemize}
We must also warn the reader that the solutions are found by solving the complicated integro-differential equation shown in Eq.~\eqref{eqvec}. As such, the exact features and origins cannot be determined more precisely. How the complicated angular dependence, scattering strengths, quadratic-in-momentum terms, conservation of net electric charge, etc. feed into the equations to bring out the final features can only be obtained via the actual numerical solutions.


\section{Comparison with results obtained from relaxation-time approximation}
\label{secsig}

In this section, we employ the simplistic application of the RTA, which we have used in our earlier works \cite{ips_rahul_ph_strain, rahul-jpcm, ips-ruiz, ips-rsw-ph, ips-shreya, ips-tilted, ips-internode, ips-spin1-ph}.
In this approach, the intranode- and internode-scattering-induced currents need to be computed separately, with the multifold nature of the nodal points to be accounted for while calculating the internodal parts \cite{ips-internode}. In order to obtain closed-form analytical expressions, we have to expand the $B $-dependent terms upto a given order in $B$, assuming it has a small magnitude, which is anyway required to justify neglecting the formation of the Landau levels. With this in mind, the various quantities are expanded as follows:
\begin{align}
f_0^\prime \left(\xi _{\chi ,s}\right) = f_0^\prime \left( \varepsilon_{\chi ,s}\right)
+ \varepsilon_{\chi ,s}^{(m)} \, f_0^{\prime \prime } \big ( \varepsilon_{\chi,s} \big )
+ \frac{1}{2} \left[ \varepsilon_{\chi ,s}^{(m)} \right ]^2 
f_0^{\prime \prime \prime}\big ( \varepsilon_{\chi ,s} \big ) + \mathcal{O} (B^3 )\,,
\end{align}
and
\begin{align}
\mathcal{D}_{\chi, s} = 
\sum \limits_{n=0}^{2} 
\left [ -e \,  \Omega_{\chi, s}^z \, B \right ]^n 
+ \mathcal{O} (B^3 )\,.
\end{align}
Here, the ``prime'' superscript denotes the operation of partial-differentiation, with respect to the variable shown explicitly within the brackets [e.g., $ f_0^\prime (\varepsilon) \equiv \partial_\varepsilon f_0 (\varepsilon)$]. 
Since we are working in the $T \rightarrow 0 $ limit, we have to use $f_0^\prime (\varepsilon) \rightarrow -\, 
\delta ( \varepsilon - \mu )$. Observing that the radial part of the integrals is with respect to the variable $k$, we need to use the following forms of the concerned expressions:
\begin{align}
& f_0^{\prime \prime } ( \varepsilon_{\chi ,s} (k) ) 
= \frac{ \partial_k f_0^\prime ( \varepsilon_{\chi ,s} (k) ) }
{2 \, c \, k \, v_F + s \, v_F} \,, \quad
f_0^{\prime \prime \prime} ( \varepsilon_{\chi ,s} (k) ) 
= \frac{ \partial^2_k f_0^\prime ( \varepsilon_{\chi ,s} (k) )
}  {v_F^2 \, (2 \, c\, k + s)^2 }
-
\frac{2 \,c\, \partial_k f_0^\prime ( \varepsilon_{\chi ,s} (k) ) 
}  {v_F^2 \, (2 \, c\, k + s)^3} \,.
\end{align}

For the intranode parts, the analytical expressions are evaluated using the mother formula of
\cite{ips-rsw-ph, ips-spin1-ph}
\begin{align}
\label{eqsig}
   \sigma^{\rm intra}_{i j} (\chi, s)
&= - \,e^2 \, \tau  
\int \frac{ d^3 \boldsymbol k}{(2\, \pi)^3 } \, \mathcal{D}_{\chi,s} 
\left[  (w_{\chi,s})_i \, + (W_\chi^s)_i \right ]
\left [ (w_{\chi,s})_j \, + (W_\chi^s)_j \right] \,
f^\prime_0 \big (\xi_{\chi,s} \big )  \,,
\text{ where }
\boldsymbol{W}_\chi^s 
= e \left  ( {\boldsymbol w}_{\chi,s} \cdot 
  \boldsymbol {\Omega}_{\chi,s} \right  ) \boldsymbol{B}\,.
\end{align}
Using the results from Ref.~\cite{ips-internode}, the conductivity for the internode-scattering-induced current is captured by
\begin{align}
\label{eqiso}
& \sigma^{\rm inter }_{ij} (\chi, s) = 
\frac {e^2  
\left [  \tau_{\text{\tiny{inter}}}  \, \rho_{-\chi}^{(0)} 
- \tau \, \rho_G^{(0)} \right ]
 \mathcal{Z}_{\chi, i}^s \, \zeta^\chi_j   }
 { \rho_G^{(0)} \, \rho_\chi^{(0)} } 
 + \order{B^3} \,,
\nn & \rho_\chi^{(0)} =
 \sum \limits_{ \tilde s }
\int \frac{ d^3 \boldsymbol q}  {(2\, \pi)^3 } 
 \left \lbrace  - f_0^\prime 
 \big (\varepsilon_{  \chi, \tilde s }  (q) \big)  \right \rbrace ,\quad
\rho_G^{(0)} = 
\frac{ \rho_\chi^{(0)} + \rho_{-\chi}^{(0)} }  {2}   \,, 
\quad \zeta^\chi_j  = 
 \sum_{\tilde s}  \mathcal{Z}_{\chi, j }^{\tilde s}  \,,
\nn &  \mathcal{Z}_{\chi, j}^s
=   B_j \int
\frac{ d^3 \boldsymbol k} {(2\, \pi)^3 } 
\left [
 e \, {\boldsymbol \Omega}_{\chi,s} (\boldsymbol k) 
 \cdot  {\boldsymbol   v}_{\chi, s} (\boldsymbol k) 
\left \lbrace - 
 f_0^\prime \big (\varepsilon_{\chi, s} (k) \big)  
 \right  \rbrace
+
  \left ( m_{\chi, s} (\boldsymbol k) \right)_j  
  \left (  v_{\chi ,s} (\boldsymbol k) \right)_j
f_0^{\prime \prime} \big (\varepsilon_{\chi,s} (k)  \big) 
\right ] ,
\end{align}
which is applicable for our isotropic bands. Here, $  \tau_{\text{\tiny{inter}}}  $ represents the relaxation time for internode scatterings, which is taken to be the same for all concerned bands. For the special case where we have scatterings between two nodes of the same nature, with no net energy-offset between the nodal points (relative to each other in the BZ), Eq.~\eqref{eqiso} further simplifies to~\cite{ips-internode}
\begin{align}
\label{eqsamenode}
 \sigma^{\rm inter }_{ij} (\chi, s)
 & =
 \frac { e^2   \, \tau  }{ \rho_1^{(0)} } 
  \left(\frac{ \tau_{\text{\tiny{inter}}} }{\tau }-1\right)
 \mathcal{Z}^{s}_{1,i} \, \sum \limits_{\tilde s} \mathcal{Z}^{\tilde s}_{1,j}\,.
\end{align}

For the configuration considered here, only the $\sigma^{\rm intra}_{zz} (\chi, s)$- and $\sigma^{\rm inter}_{zz} (\chi, s)$-components are nonzero. Moreover, due to the isotropic nature of the dispersions, only even powers of $B$ can appear, as also justified by the Onsager-Casimir reciprocity relation, $\sigma^{\rm intra(inter)}_{zz} (\mathbf{B})= \sigma^{\rm intra(inter)}_{zz} (-\mathbf{B})$ \cite{onsager31_reciprocal, onsager2, onsager3}. Consequently, the answers turn out to be $\chi$-independent.

\subsection{Contribution from intranode parts}

For the $s=1$ and $s=0$ bands, Eq.~\eqref{eqsig} evaluates to
\begin{align}
&  \sigma_{zz}^{\rm dr} \big \vert_{s=1} =
\frac{e^2\, \tau \, \sqrt{ 4 \, c \, \mu  + v_F } 
\left(\sqrt{v_F}-\sqrt{4 \, c \, \mu +v_F}\right)^2} 
{(2\,\pi)^3 \times 6 \, c^2 \, \sqrt v_F}
\,, \quad
\sigma_{zz}^{\rm bc} \big \vert_{s=1} =
\frac{64 \,B^2 \,c^2 \,e^4 \,\tau \, v_F^{\frac{3} {2} } 
\;\sqrt{4 \,c \, \mu + v_F}}
{(2\,\pi)^3 \times  15 \left(\sqrt{v_F}-\sqrt{4\, c \,\mu +v_F}\right)^2}\,,
\nn & \sigma_{zz}^{\rm omm} \big \vert_{s=1} =
\frac{ - \,2 \,B^2 \,c^2 \,e^4 \,\tau  \,v_F^{\frac{3} {2} } 
\left[  12  \,v_F^2
+ 42 \, c  \,\mu  \, v_F + 16  \,c \, \mu  \,
 \sqrt{ v_F \left(4  \,c  \,\mu + v_F\right)}
 + \sqrt{ v_F^3
   \left(4  \, c  \,\mu +v_F\right)} \right ]}
   {(2\,\pi)^3 \times  15 \left(4  \,c  \,\mu +v_F\right)^{\frac{3} {2} } \left(\sqrt{v_F}
  -\sqrt{4  \, c  \, \mu  + v_F} \right )^2} \,,
\end{align}
\begin{align}
&  \sigma_{zz}^{\rm dr} \big \vert_{s=0} = 
\frac{4 \,e^2\, \mu ^{\frac{3} {2} } \,\tau }  {(2\,\pi)^3 \times  3 \, \sqrt{c \,v_F} } \,,\quad
\sigma_{zz}^{\rm bc} \big \vert_{s=0} = 0\,,\quad
\sigma_{zz}^{\rm omm} \big \vert_{s=0} =
\frac{7\, B^2\, \sqrt{c} \,e^4 \,\tau \, v_F^{5/2}}
{ (2\,\pi)^3 \times  30 \, \mu ^{\frac{3} {2} }}\,.
\end{align}
Here, the superscipts, ``dr'', ``bc'', and ``omm'', indicate the Drude (i.e., $B$-independent), BC-only, and OMM-induced contributions, respectively. For $s=1$, $\sigma_{zz}^{\rm bc} $ and $\sigma_{zz}^{\rm omm}$ come with opposite signs. Within the parameter regions for $\mu$, $v_F$, and $c$ considered here, the magnitude of the latter is always smaller than that of the former, with the overall response remaining positive. This is in agreement with the conclusions reached in Ref.~\cite{ips-spin1-ph}, where the quadratic-in-momentum corrections were not considered. For the $s=0$ bands, the $B$-dependent part is solely sourced by the OMM and the relevant terms involve a product of two terms arising from OMM. Such a product term gives only a positive contribution, as is reflected in $\sigma_{zz}^{\rm omm} \big \vert_{s=0} $. In fact, the negative parts in $\sigma_{zz}^{\rm omm} \big \vert_{s=1}$ have arisen from integrands containing a product of one power of BC-only and one power of OMM-only terms, cancelling out (and overpowering) the OMM-only positive terms. From all these discussions, the conclusion is that, $  \sigma_{zz}^{\rm intra} = \sigma_{zz}^{\rm bc} + \sigma_{zz}^{\rm omm} $ turns out to be exclusively positive for all the bands within the RTA, without any possibility of sign-flips.

\subsection{Contribution from internode parts}

Evaluating Eq.~\eqref{eqsamenode} for our system, we obtain
\begin{align}
 \sigma_{zz}^{\rm inter} \big \vert_{s=1} & =
\frac{ B^2 \, e^4 \, \tau  \left(c \,  v_F\right)^{\frac{3} {2} }
 \sqrt{4 \,c \,\mu + v_F}
  \left(\sqrt{\frac{v_F} {4 \, c \,\mu +v_F}}+6\right) 
  \left [ \sqrt{c\, \mu \, v_F}
   \left ( \sqrt{\frac{v_F}{4 \, c\, \mu +v_F}}+6\right) + v_F\right ]  
 } 
 {36 \, \pi ^2\,\sqrt \mu 
 \left(  v_F^{\frac{3} {2} } - v_F \, \sqrt{4 \, c \, \mu + v_F}
  + 2 \, c \, \mu\, \sqrt{v_F} + \sqrt{c \,\mu\, v_F} 
  \, \sqrt{4 \, c \, \mu + v_F} \right)}
 \left(\frac{ \tau_{\text{\tiny{inter}}} }{\tau }-1\right)  ,\nn
 \sigma_{zz}^{\rm inter} \big \vert_{s=0} & =
 \frac{B^2 \,c\, e^4 \, \tau \, v_F^2 
 \sqrt{ 4\, c \, \mu  +v_F}}{36 \, \pi ^2 \, \mu 
  \left( v_F^{\frac{3} {2} } -v_F \sqrt{ 4\, c \, \mu  +v_F} 
  + 2 \, c  \, \mu \,  \sqrt{v_F}
  +\sqrt{c \,\mu \, v_F} \,\sqrt{ 4\, c \, \mu  +v_F}
  \right)}
 \left(\frac{ \tau_{\text{\tiny{inter}}} }{\tau }-1\right) .
\end{align} 
In realistic systems (see, for example, Ref.~\cite{claudia-multifold}), $ { \tau_{\text{\tiny{inter}}} }/{\tau }  \gg 1$, leading to
$\sigma_{zz}^{\rm inter}$ being positive for each band, and with a magnitude much much larger than $\sigma_{zz}^{\rm intra}$.

\subsection{Overall characteristics}

The net conductivity from each band turns out to be positive, with the internode part being the dominating contribution.
Hence, we find that the RTA is unable to capture the actual characteristics that have been obtained by the exact solutions of the Boltzmann equations. Albeit, the results obtained by the former can be improved upon by introducing interband scatterings within a single node and, perhaps, by introducing band-dependent phenomenological values of the relaxation times.

\section{Summary, discussions, and future perspectives}
\label{secsum}

In this paper, we have addressed the crucial question of determining the nature of longitudinal magnetoconductivity for multifold semimetals, taking the TSM with quadratic-corrections as an example. A TSM's threefold-degenerate nodal points ensure that multiple bands contribute to transport at each node, although this signature was missed/ignored in earlier works \cite{ips-spin1-ph, girish-internode-spin1}, because the $s=0$ band provided zero input due its designation as a flat/nondispersive band. However, the flatness of this band is an artefact of retaining only the linear-in-momentum terms in the effective Hamiltonian in the vicinity of a given node, which is corrected in our approach by including the leading-order corrections $\propto k^2$. Indeed, the $s=0$ bands harbour nontrivial topological characteristics on the virtue of possessing nonzero values of OMM, although having zero BC (or vanishing Chern numbers). Hence, the $c\, k^2$ term in the Hamiltonian [cf. Eq.~\eqref{eqham}] enables us to account for the actual response that must be observable in contemporary experiments \cite{claudia-multifold}. Our calculations and results also establish the importance of going beyond the RTA, as the full/correct picture is not obtained by confining ourselves to such an approximation. This is explicitly and quantitatively verified by computing and comparing the answers obtained from the two approaches. Indeed, the failure of an RTA has been observed in diverse contexts (see, for example Refs. \cite{rta1, rta2}), where the correct physics is only captured by an exact treatment of the collision integrals.

Our results show that a nonzero longitudinal conductivity in TSMs is caused by chiral anomaly, similar to the WSMs. The inclusion of OMM also leads to sign reversals of $\delta \sigma_{zz}$ depending on the relative strengths of internode- and intranode-scatterings, similar to the WSMs. The qualitative difference from the WSM-cases is that, since more than one band is participating in transport at each chiral node, we now have multiple knobs available to tune the characteristics of the linear response. These comprise various types of intraband- and intranode-scattering-amplitudes, viz. the set comprising $ \lbrace \beta^{\rm intra}_{s, \tilde s}, \,\beta^{\rm inter}_{s, \tilde s} \rbrace $. Quantitatively, we find that a large ratio of internode-to-intranode scattering-strength is required (cf. Fig.~\ref{figall3}) to get the sign-flips of $\delta \sigma_{zz}$, compared to WSMs \cite{timm}.

To summarise the importance of our work, we emphasise on the following points: This is the first paper to incorporate a proper quadratic correction in the so-called flat-band of the TSMs while computing conductivity. This situation is highly nontrivial because the flat-band possesses a nontrivial OMM, despite having vanishing BC. In all prior literature \cite{pal22b_berry, ips-spin1-ph, ips-internode, girish-internode-spin1}, the conductivity from the flat-band has been omitted since, at leading order, it has zero dispersion. But this omission is not physical, as explained earlier. As such, our analysis here represents a significant advancement in the study of multifold semimetals in 3d. Secondly, the technique of solving the Boltzmann’s equations exactly (i.e., by going beyond the RTA) \cite{timm}, when more than one band is participating in transport at each chiral node, has not been implemented earlier. Our study of TSMs allows exactly that, demonstrating how intraband and interband scattering amplitudes can tune inear response. Last but not the least, it is a nontrivial task to extract the analytical expressions even within the RTA. We do that as well and bring out the inadequacies in the results obtained from RTA. Our investigations are very timely, thanks to the ongoing experimental efforts to probe novel transport properties of multifold fermions \cite{claudia-multifold}.

In the future, it will be worthwhile to repeat our calculations for a magnetic field which is not exactly collinear with the electric field \cite{girish2023}. This will cause nonzero components of conductivity in the form of Hall and planar-Hall response. Another avenue to explore is to introduce nonzero tilts in the spectra \cite{das-agarwal_omm, rahul-jpcm, ips-tilted, ips-shreya, ips_tilted_dirac}. Since realistic bandstructures indeed show tilting, this constitutes an important aspect. In particular, tilting can cause linear-in-$B$ terms to appear in the conductivity~\cite{rahul-jpcm, ips-tilted, das-agarwal_omm, ips-tilted, ips-shreya, girish2023, ips_tilted_dirac}, satisfying Onsager-Casimir reciprocity relations. Next, one would like to determine the characteristics of magneto-optical conductivity under quantising magnetic fields~\cite{gusynin06_magneto, staalhammar20_magneto, yadav23_magneto} for the TSMs. A straightforward but interesting complementary direction is to compute transmission charactertistics in tunneling problems in the quadratic-corrected TSMs, in the same spirit as done in some of our earlier works \cite{fang, zhu, ips3by2, deng2020, ips-aritra, ips-jns}. A nonlinear-in-momentum dependence in the propagation direction of the quasiparticles brings in the existence of evanescent waves, which render the problem computationally challenging to solve \cite{banerjee, ips-abs-semid, deng2020, ips-aritra, ips-jns, ips_tunnel_qbcp_corr, ips_tunnel_qbcp_delta}.

\section*{Acknowledgments}
We thank Carsten Timm for useful discussions. This research, leading to the results reported, has received funding from the Council of Science \& Technology (CST), U.P.

\section*{Data availability}

All data generated or analysed during this study are included within the article.

\appendix

\section{Matrix equation for determining the unknown coefficients}
\label{appmat}

The two matrices, $\mathcal A$ and $\mathcal H$, shown in Eq.~\eqref{eqmatrix}, are defined by
{\tiny
\begin{align*}
& - \left[ 
\begin{matrix}
c_{11}^8 \, \beta_{1,1}^{\text{intra}} -1
& \text{cs}_{11}^{44} \, \beta_{11}^{\text{intra}}
& \text{cs}_{11}^{42} \, \beta_{1,1}^{\text{intra}} 
 & \text{cs}_{21}^{44} \, \beta_{1,1}^{\text{inter}} 
   & s_{21}^8 \, \beta_{1,1}^{\text{inter}} \\ \\
 \text{cs}_{11}^{44} \, \beta_{1,1}^{\text{intra}} 
 & s_{11}^8 \, \beta_{1,1}^{\text{intra}} -1
 & \text{ss}_{11}^{42} \, \beta_{1,1}^{\text{intra}} 
 & c_{21}^8 \, \beta_{1,1}^{\text{inter}} 
   & \text{cs}_{21}^{44} \, \beta_{1,1}^{\text{inter}} \\ \\
\frac{ \text{cs}_{11}^{42} \, \beta_{1,1}^{\text{intra}} }{4} 
 & \frac{  \text{ss}_{11}^{42} \, \beta_{1,1}^{\text{intra}} }{4} 
  &\frac{  s_{11}^4 \, \beta_{11}^{\text{intra}}- 4}   {4} 
   & \frac{  \text{cs}_{21}^{42} \, \beta_{1,1}^{\text{inter}} }{4} 
   & \frac{  \text{ss}_{21}^{42} \, \beta_{1,1}^{\text{inter}} }{4} \\ \\
\text{cs}_{11}^{44} \, \beta_{1,1}^{\text{inter}} 
& s_{11}^8 \, \beta_{1,1}^{\text{inter}} 
 & \text{ss}_{11}^{42} \, \beta_{1,1}^{\text{inter}} 
 &  c_{21}^8 \, \beta_{1,1}^{\text{intra}} -1
   & \text{cs}_{21}^{44} \, \beta_{1,1}^{\text{intra}} \\ \\
 c_{11}^8 \, \beta_{1,1}^{\text{inter}} 
 & \text{cs}_{11}^{44} \, \beta_{11}^{\text{inter}}
  & \text{cs}_{11}^{42} \, \beta_{1,1}^{\text{inter}} 
 & \text{cs}_{21}^{44} \, \beta_{1,1}^{\text{intra}} 
& s_{21}^8 \, \beta_{1,1}^{\text{intra}} - 1 \\ \\
 \frac{  \text{cs}_{11}^{42} \, \beta_{1,1}^{\text{inter}} }{4} 
 & \frac{  \text{ss}_{11}^{42} \, \beta_{1,1}^{\text{inter}} }{4} 
 & \frac{  s_{11}^4 \, \beta_{1,1}^{\text{inter}} }{4} 
 & \frac{  \text{cs}_{21}^{42} \, \beta_{1,1}^{\text{intra}} }{4} 
 & \frac{ \text{ss}_{21}^{42} \, \beta_{1,1}^{\text{intra}} }{4}  \\ \\
 \frac{ \left(c_{11}^8+\text{cs}_{11}^{44} \right)  \beta_{1,0}^{\text{intra}} }{2} 
 & \frac{ \left(\text{cs}_{11}^{44}+s_{11}^8\right) \beta_{1,0}^{\text{intra}} }{2} 
& \frac{ \left(\text{cs}_{11}^{42}+\text{ss}_{11}^{42} \right)  \beta_{1,0}^{\text{intra}} }{2} 
   & \frac{  \left(c_{21}^8+\text{cs}_{21}^{44} \right)  \beta_{1,0}^{\text{inter}} }{2} 
   & \frac{ \left(\text{cs}_{21}^{44}+s_{21}^8 \right)  \beta_{1,0}^{\text{inter}} }{2}  \\ \\
 \frac{ \text{cs}_{11}^{42} \, \beta_{1,0}^{\text{intra}} }{2} 
 & \frac{ \text{ss}_{11}^{42} \, \beta_{1,0}^{\text{intra}} }{2} 
  & \frac{s_{11}^4 \, \beta_{1,0}^{\text{intra}} }{2} 
  & \frac{\text{cs}_{21}^{42} \, \beta_{1,0}^{\text{inter}} }{2} 
   & \frac{\text{ss}_{21}^{42} \, \beta_{1,0}^{\text{inter}} }{2}  \\ \\
 \frac{ \left(c_{11}^8+\text{cs}_{11}^{44} \right)  \beta_{1,0}^{\text{inter}} }{2} 
 & \frac{ \left(\text{cs}_{11}^{44}+s_{11}^8\right)\, \beta_{1,0}^{\text{inter}} }{2} 
& \frac{ \left(\text{cs}_{11}^{42}+\text{ss}_{11}^{42} \right)  \beta_{1,0}^{\text{inter}} }{2} 
   & \frac{\left(c_{21}^8 + \text{cs}_{21}^{44} \right)  \beta_{1,0}^{\text{intra}} }{2} 
   & \frac{\left(\text{cs}_{21}^{44}+s_{21}^8 \right)  \beta_{10}^{\text{intra}}}{2}   \\ \\
 \frac{ \text{cs}_{11}^{42} \, \beta_{1,0}^{\text{inter}} }{2}
 & \frac{\text{ss}_{11}^{42} \, \beta_{10}^{\text{inter}}}{2} 
  & \frac{s_{11}^4 \, \beta_{1,0}^{\text{inter}} }{2} 
  & \frac{ \text{cs}_{21}^{42} \, \beta_{1,0}^{\text{intra}} }{2} 
   & \frac{ \text{ss}_{21}^{42} \, \beta_{1,0}^{\text{intra}} }{2} \\\\
\end{matrix}\right.\nn
& \hspace{ 6 cm}
\left.\begin{matrix}   
 \text{ss}_{21}^{42} \,\, \beta_{11}^{\text{inter}} 
   & \frac{  s_{10}^4 \, \beta_{1,0}^{\text{intra}} }{2}
   & \frac{  \text{cs}_{10}^{22} \, \beta_{1,0}^{\text{intra}} }{2}
   & \frac{  s_{20}^4 \, \beta_{1,0}^{\text{inter}} }{2}
   & \frac{ \text{cs}_{20}^{22} \, \beta_{1,0}^{\text{inter}} }{2}\\ \\
 \text{cs}_{21}^{42} \,\beta_{1,1}^{\text{inter}} 
   & \frac{ s_{10}^4 \, \beta_{1,0}^{\text{intra}} }{2}
   & \frac{ \text{cs}_{10}^{22} \, \beta_{1,0}^{\text{intra}} }{2}
   & \frac{ s_{20}^4 \, \beta_{1,0}^{\text{inter}} }{2}
   & \frac{ \text{cs}_{20}^{22} \, \beta_{1,0}^{\text{inter}} }{2}\\ \\
\frac{s_{21}^4 \, \beta_{1,1}^{\text{inter}} }{4} 
   & \frac{ \text{cs}_{10}^{22} \, \beta_{1,0}^{\text{intra}} }{2}
   & \frac{  c_{10}^4 \, \beta_{1,0}^{\text{intra}} }{2}
   & \frac{  \text{cs}_{20}^{22}\, \beta_{1,0}^{\text{inter}} }{2}
   & \frac{ c_{20}^4 \, \beta_{1,0}^{\text{inter}} }{2}\\
\text{cs}_{21}^{42} \,\beta_{11}^{\text{intra}}
    & \frac{  s_{10}^4 \, \beta_{1,0}^{\text{inter}} }{2}
    & \frac{  \text{cs}_{10}^{22} \, \beta_{1,0}^{\text{inter}} }{2}
   & \frac{  s_{20}^4 \, \beta_{1,0}^{\text{intra}} }{2}
   & \frac{ \text{cs}_{20}^{22} \, \beta_{1,0}^{\text{intra}} }{2} \\ \\
  \text{ss}_{21}^{42} \,\, \beta_{11}^{\text{intra}} 
   & \frac{ s_{10}^4 \, \beta_{1,0}^{\text{inter}} }{2} 
   & \frac{ \text{cs}_{10}^{22} \, \beta_{10}^{\text{inter}}}{2} 
   & \frac{ s_{20}^4 \, \beta_{1,0}^{\text{intra}} }{2} 
   & \frac{ \text{cs}_{20}^{22} \, \beta_{1,0}^{\text{intra}} }{2} \\  \\
  \frac{  s_{21}^4 \, \beta_{1,1}^{\text{intra}} -4}  {4} 
   & \frac{\text{cs}_{10}^{22} \, \beta_{1,0}^{\text{inter}} }{2} 
   & \frac{ c_{10}^4 \, \beta_{1,0}^{\text{inter}} }{2} 
   & \frac{ \text{cs}_{20}^{22} \, \beta_{1,0}^{\text{intra}} }{2} 
   & \frac{ c_{20}^4 \, \beta_{1,0}^{\text{intra}} }{2} \\ \\
 \frac{  \left(\text{cs}_{21}^{42}+\text{ss}_{21}^{42} \right)  \beta_{1,0}^{\text{inter}} }{2} 
   & \frac{  s_{10}^4 \, \beta_{0,0}^{\text{intra}} -2} {2}
   & \frac{  \text{cs}_{10}^{22} \, \beta_{0,0}^{\text{intra}} }{2} 
   & \frac{ s_{20}^4 \, \beta_{0,0}^{\text{inter}} }{2} 
   & \frac{  \text{cs}_{20}^{22} \, \beta_{00}^{\text{inter}}}{2}  \\ \\
 \frac{ s_{21}^4 \, \beta_{1,0}^{\text{inter}} }{2} 
   & \text{cs}_{10}^{22} \,\beta_{0,0}^{\text{intra}} 
   & c_{10}^4 \, \beta_{0,0}^{\text{intra}} - 1
   & \text{cs}_{20}^{22} \, \beta_{0,0}^{\text{inter}} 
   & c_{20}^4 \,\beta_{0,0}^{\text{inter}} \\ \\
  \frac{\left(\text{cs}_{21}^{42}+\text{ss}_{21}^{42} \right)  \beta_{1,0}^{\text{intra}} }{2} 
   & \frac{ s_{10}^4 \, \beta_{0,0}^{\text{inter}} }{2} 
   & \frac{ \text{cs}_{10}^{22} \, \beta_{0,0}^{\text{inter}} }{2} 
   & \frac{ s_{20}^4 \, \beta_{0,0}^{\text{intra}} -2} {2}
   & \frac{\text{cs}_{20}^{22} \, \beta_{00}^{\text{intra}}}{2}  \\ \\
\frac{ s_{21}^4 \, \beta_{1,0}^{\text{intra}} } {2}
   & \text{cs}_{10}^{22} \,\beta_{0,0}^{\text{inter}} 
   & c_{10}^4 \, \beta_{0,0}^{\text{inter}} 
   & \text{cs}_{20}^{22} \, \beta_{0,0}^{\text{intra}} 
  & c_{20}^4 \,\beta_{0,0}^{\text{intra}} -1 \\ \\
\end{matrix} 
\right ]
\end{align*}
}
and
{\tiny
\begin{align*}
\frac{1}{2}   \begin{bmatrix}
 2\, \text{hc}_{11}^4 \, \beta_{1,1}^{\text{intra}}+ \text{hs}_{20}^2 \, \beta_{1,0}^{\text{inter}}
 +2\, \text{hs}_{21}^4 \,\beta_{1,1}^{\text{inter}}+ \text{hs}_{10}^2 \, \beta_{1,0}^{\text{intra}} \\ \\
 2\, \text{hc}_{21}^4 \, \beta_{1,1}^{\text{inter}}+ \text{hs}_{20}^2 \, \beta_{1,0}^{\text{inter}}+ \text{hs}_{10}^2 
 \,\beta_{1,0}^{\text{intra}}+2\, \text{hs}_{11}^4 \, \beta_{1,1}^{\text{intra}} \\ \\
 \frac{\ 2\, \text{hc}_{20}^2 \, \beta_{1,0}^{\text{inter}}
 +2\, \text{hc}_{10}^2 \, \beta_{1,0}^{\text{intra}}
 + \text{hs}_{21}^2\, \beta_{1,1}^{\text{inter}}+ \text{hs}_{11}^2 \, \beta_{1,1}^{\text{intra}}
 } {2} \\ \\
 2\, \text{hc}_{21}^4 \, \beta_{1,1}^{\text{intra}}
 + \text{hs}_{10}^2 \, \beta_{1,0}^{\text{inter}}+2\, \text{hs}_{11}^4 \beta
   _{1,1}^{\text{inter}}+ \text{hs}_{20}^2 \, \beta_{1,0}^{\text{intra}} \\ \\
 2\, \text{hc}_{11}^4 \, \beta_{1,1}^{\text{inter}}+ \text{hs}_{10}^2 \, \beta_{1,0}^{\text{inter}}
 + \text{hs}_{20}^2 \, \beta_{1,0}^{\text{intra}}+2\, \text{hs}_{21}^4 \, \beta_{1,1}^{\text{intra}} \\ \\
 \frac{2\, \text{hc}_{10}^2 \, \beta_{1,0}^{\text{inter}}
 +2\, \text{hc}_{20}^2 \, \beta_{1,0}^{\text{intra}}
 + \text{hs}_{11}^2
   \, \beta_{1,1}^{\text{inter}}+ \text{hs}_{21}^2 \, \beta_{1,1}^{\text{intra}}
}{2}    \\ \\
 \left( \text{hc}_{21}^4+ \text{hs}_{21}^4\right) \, \beta_{1,0}^{\text{inter}}
 +\left( \text{hc}_{11}^4+ \text{hs}_{11}^4\right) 
 \,\beta_{1,0}^{\text{intra}}+ \text{hs}_{20}^2 \, \beta_{0,0}^{\text{inter}}
 + \text{hs}_{10}^2 \, \beta_{0,0}^{\text{intra}} \\ \\
 2\, \text{hc}_{20}^2 \, \beta_{0,0}^{\text{inter}}
 +2\, \text{hc}_{10}^2 \, \beta_{0,0}^{\text{intra}}+ \text{hs}_{21}^2 
 \,\beta_{1,0}^{\text{inter}}+ \text{hs}_{11}^2 \, \beta_{1,0}^{\text{intra}} \\ \\
 \left( \text{hc}_{11}^4+ \text{hs}_{11}^4\right) \, \beta_{1,0}^{\text{inter}}
 +\left( \text{hc}_{21}^4+ \text{hs}_{21}^4\right) \beta_{1,0}^{\text{intra}}+ \text{hs}_{10}^2 \, \beta_{0,0}^{\text{inter}}
 + \text{hs}_{20}^2 \, \beta_{0,0}^{\text{intra}} \\ \\
 2\, \text{hc}_{10}^2 \, \beta_{0,0}^{\text{inter}}+2\, \text{hc}_{20}^2 \, \beta_{0,0}^{\text{intra}}
 + \text{hs}_{11}^2
 \, \beta_{1,0}^{\text{inter}}+ \text{hs}_{21}^2 \, \beta_{1,0}^{\text{intra}} \\
\end{bmatrix},
\end{align*}
}
respectively.
The various symbols represent the following integrals:
\begin{align}
& c_{\alpha_\chi s}^4 = \int d\theta \, {\mathcal F}_{\chi, s} (k,\theta)  \,\cos^4 \theta  \,,\quad
s_{\alpha_\chi s}^4 = \int d\theta \, {\mathcal F}_{\chi, s}(k,\theta)  \, \sin^4 \theta \,, \quad
\text{cs}_{\alpha_\chi s}^{22} = \int d\theta \, {\mathcal F}_{\chi, s} (k,\theta)  \,\cos^2 \theta \sin^2  \theta \,,\nn
&
\text{hc}_{\alpha_\chi s}^2 = \int d\theta \, {\mathcal F}_{\chi, s} (k,\theta)  \,\cos^2 \theta 
\, h_{\chi, s} (\mu, \theta) \,,\quad
\text{hs}_{\alpha_\chi s}^2 = \int d\theta \, {\mathcal F}_{\chi, s}(k,\theta)  \, \sin^2 \theta \, 
h_{\chi, s} (\mu, \theta)\,,\nn &
 \text{hc}_{\alpha_\chi s}^4 = \int d\theta \, {\mathcal F}_{\chi, s} (k,\theta)  \,\cos^4 ( \theta/2 ) 
\, h_{\chi, s} (\mu, \theta)\,, \quad
\text{hs}_{\alpha_\chi s}^4 = \int d\theta \, {\mathcal F}_{\chi, s}(k,\theta)  \, 
\sin^4  ( \theta/2 )\, h_{\chi, s} (\mu, \theta) \,,\nn &
c_{\alpha_\chi s}^8 = \int d\theta \, {\mathcal F}_{\chi, s} (k,\theta)  \,\cos^8 ( \theta/2 ) \,,\quad
s_{\alpha_\chi s}^8 = \int d\theta \, {\mathcal F}_{\chi, s}(k,\theta)  \, \sin^8  ( \theta/2 ) \,,
\quad \text{cs}_{\alpha_\chi s}^{44} = \int d\theta \, {\mathcal F}_{\chi, s} (k,\theta)  
\,\cos^4 ( \theta/2 ) \sin^4 ( \theta/2 ) \,,
\nn &\text{cs}_{\alpha_\chi s}^{42} = \int d\theta \, {\mathcal F}_{\chi, s} (k,\theta)  \,
\cos^4 ( \theta/2 ) \sin^2  \theta \,,\quad
\text{ss}_{\alpha_\chi s}^{42} = \int d\theta \, {\mathcal F}_{\chi, s} (k,\theta)  \,\sin^4 ( \theta/2 ) \sin^2  \theta \,,
\end{align}
where
\begin{align}
\alpha_\chi = \begin{cases}
1 &\text{ for } \chi =+1 \\
2 &\text{ for } \chi = -1
\end{cases},
\end{align}
and
\begin{align}
{\mathcal F}_{\chi, s} (\mu,\theta) = \tau_{\chi, s} (\mu,\theta)\,
 \frac{\sin \theta \,k^3 
\, {\mathcal D}^{-1}_{ \chi,   s} (k, \theta)
}
{  |\boldsymbol k  \cdot {\boldsymbol{w}}_{\chi, s} (\boldsymbol k)  | }
\Bigg \vert_{ k  = k_F^{ \chi, s} (\theta) } \,.
\end{align}

\section{The logic behind determing the ansatz for ${\Lambda}^z_\chi (\mu,\theta )$}
\label{appansatz}

For the benefit of the reader, we explain here the logic behind determining the ansatz for the function ${\Lambda}^z_\chi (\mu,\theta )$. Basically, one must ensure that $ {\Lambda}^z_{\chi, s} (\mu,\theta ) \,  {\Lambda}^z_{\tilde \chi, \tilde s} (\mu,\theta^\prime )$ contains all sinusoidal terms appearing in 
\begin{align}
{\mathcal T }^{\chi, \tilde \chi}_{ s, \tilde s} (\theta, \theta^\prime)
& = \left [
\sin ^4\bigg(\frac{\theta }{2}\bigg) \, \sin ^4\bigg (\frac{\theta '}{2}\bigg )
+\frac{1}{4} \sin ^2 \theta  \, \sin ^2 \theta'
+ \cos ^4\bigg(\frac{\theta }{2}\bigg ) \cos^4\bigg (\frac{\theta '}{2}\bigg )
\right ] \delta_{s,1} \, \delta_{\tilde s, 1}
 \nn & \quad 
+ \left [
\frac{ \sin ^2 \theta \, \sin ^2 \theta ' }{2} 
+\cos ^2 \theta \, \cos ^2 \theta '
\right ] \delta_{s,0} \, \delta_{\tilde s, 0}
\nn & \quad + \left [ 
\left\lbrace \sin ^4\bigg (\frac{\theta }{2}\bigg )
+ \cos ^4\bigg( \frac{\theta }{2}\bigg )
\right \rbrace \sin ^2 \theta '
+\sin ^2 \theta \, \cos ^2\theta '  \right ]   \delta_{s,1} \, \delta_{\tilde s, 0} 
\nn & \quad + \left [ 
 \sin ^2 \theta 
\left\lbrace \sin ^4\bigg (\frac{\theta^\prime }{2}\bigg )
+ \cos ^4 \bigg( \frac{\theta ^\prime}{2}\bigg )
\right \rbrace
+\cos ^2\theta \, \sin ^2 \theta^\prime  \right ]  
\delta_{s,0} \, \delta_{\tilde s, 1} \,.
\end{align}
In fact, if one takes more terms with independent sinusoidal forms, the solutions for the corresponding coefficients are zero. For example, in Ref.~\cite{girish-internode-spin1}, the authors have included one extra spurious/unnecessary coefficient, $\lambda^\chi $ (accompanying unity or $\theta$-independent term), which identically turns out to be zero [because ${\mathcal T }^{\chi, \tilde \chi}_{ s, \tilde s} (\theta, \theta^\prime)$ does not contain any such $\theta$- or $\theta^\prime $-independent term].This has been been tested explicitly by finding the solutions. Just to demonstrate how the ansatz changes depending on the system one considers, we provide the following examples:
\begin{enumerate}
\item For a pair of chirally-conjugate Weyl nodes, analysed in Ref.~\cite{timm}, the overlap-function is
\begin{align}
{\mathcal T }_{ \chi, \tilde \chi} (\theta, \theta^\prime)
& = \frac{ 1 + \chi \, \tilde \chi \cos \theta \, \cos \theta^\prime }{2} \, .
\end{align}
Hence, the ansatz is taken to be
\begin{align}
 {\Lambda}^z_\chi (\mu,\theta ) & = 
\tau_\chi (\mu,\theta) \left [ \lambda_\chi - h_\chi (\mu, \theta) 
 + a_\chi \, \cos \theta  \right ].
\end{align}

\item For a single Kramers-Weyl node with a twofold band-crossing, analysed in Ref.~\cite{ips-exact-kwn}, the overlap-function is
\begin{align}
{\mathcal T }_{ s, \tilde s} (\theta, \theta^\prime)
& = \frac{ 1 + s \, \tilde s \cos \theta \, \cos \theta^\prime }{2} \, ,
\end{align}
for which the ansatz is taken to be
\begin{align}
 {\Lambda}^z_s (\mu,\theta ) & = 
\tau_s (\mu,\theta) \left [ \lambda_s - h_{s} (\mu, \theta) 
 + a_s \, \cos \theta  \right ].
\end{align}

\item For a single RSW node with a fourfold band-crossing, analysed in Ref.~\cite{ips-exact-rsw}, the overlap-function is
\begin{align}
{\mathcal T }_{ s, \tilde s} (\theta, \theta^\prime)
& = \left[  5 -3 \cos^2 \theta^\prime 
+ \cos  \theta  \left(17 \cos  \theta^\prime  -
27 \cos^3 \theta^\prime  \right)+\cos^2 \theta  
\left(9 \cos^2 \theta^\prime  -3\right) 
  +9 \cos^3 \theta  \cos  \theta^\prime   \left(5 \cos^2 \theta^\prime  -3\right)
    \right ] \beta_{\rm intra}^{1,1} \nn & \quad
+ \left[ 5 - 3\cos^2 \theta^\prime
+\cos  \theta  \left(9 \cos\theta^\prime  -3 \cos^3 \theta^\prime  \right)
+ \cos^2 \theta  \left(9 \cos^2 \theta^\prime  -3\right)
+ \cos^3 \theta  \left(5 \cos^3 \theta^\prime  -3 \cos  \theta^\prime  \right)  
\right ] \beta_{\rm intra}^{3,3} \nn & \quad
+ \left[ 3 + 3\cos^2 \theta^\prime  
+ \cos^3 \theta  \left(9 \cos  \theta^\prime  -15 \cos^3 \theta^\prime  \right)
+\cos  \theta  \left(9 \cos^3 \theta^\prime  -3 \cos  \theta^\prime  \right)
   +\cos^2 \theta  \left(3-9 \cos^2 \theta^\prime  \right)  
 \right ]  \beta_{\rm inter}\,.
\end{align}
Accordingly, the ansatz is taken to be
\begin{align}
 {\Lambda}^z_s (\mu,\theta ) & = 
\tau_s (\mu,\theta)
 \left [ \lambda_s - h_s + a_s \cos \theta  + b_s \cos ^2 \theta 
 + c_s \cos^3 \theta   \right ].
\end{align}
\end{enumerate}

 
\bibliography{ref_spin1}


\end{document}